\documentclass[aps,prb,twocolumn,superscriptaddress,floatfix, raggedbottom]{revtex4-2}
\usepackage[utf8]{inputenc}
\usepackage{natbib}
\usepackage[colorlinks=true,linkcolor=blue,citecolor=blue,urlcolor=blue]{hyperref}
\usepackage{graphicx}
\usepackage{latexsym}
\usepackage{amssymb}
\usepackage{amsmath}
\usepackage{amsfonts}
\usepackage{amsthm}
\usepackage{bm}
\usepackage{floatrow}
\usepackage[caption=false]{subfig}
\usepackage{bbm}
\usepackage{enumitem}
\usepackage{hyperref}
\usepackage{lipsum}
\usepackage{braket}
\usepackage{tikz}
\usepackage[margin=1in]{geometry}
\usetikzlibrary{shapes.geometric, arrows.meta}
\usepackage{pgfplots}
\usepackage{comment}
\usepackage{diagbox}
\usepackage{ifthen}
\usepackage{ragged2e}
\usepackage{multirow}
\usepackage{mathtools}
\usepackage[export]{adjustbox}
\usetikzlibrary{arrows,decorations.pathreplacing,decorations.markings,arrows.meta,patterns,3d}

\tikzset{
  midarrow/.style args={#1}{
    postaction={decorate},
    decoration={
      markings,
        mark=at position #1 with {
        \arrow{Latex[length=2.5mm,width=1.2mm]}
      }
    }
  }
}

\usepackage[normalem]{ulem} 
\usepackage{cancel}
\usepackage{microtype}
\usepackage{booktabs}

\theoremstyle{definition}
\newtheorem{proposition}{Proposition}

\newtheorem{criterion}[proposition]{Criteria}

\newcommand{\cA}{\mathcal A}
\newcommand{\cC}{\mathcal C}
\newcommand{\cL}{\mathcal L}
\newcommand{\cK}{\mathcal K}

\newcommand{\cB}{\mathcal B}

\begin{document}

\title{Field theories and critical excitations at anyon condensation transitions}
\date{\today}

\author{Luisa Eck}
\affiliation{Department of Physics and Institute for Quantum Information and Matter, \mbox{California Institute of Technology, Pasadena, CA, 91125, USA}}
\author{Xie Chen}
\affiliation{Department of Physics and Institute for Quantum Information and Matter, \mbox{California Institute of Technology, Pasadena, CA, 91125, USA}}

\begin{abstract}
Anyon condensation transitions remain challenging to characterize especially when the condensing anyons are non-abelian. We develop two complementary approaches for extracting information about the transition from the categorical data of the parent topological order and the condensate. 
The first adapts a Ginzburg-Landau procedure developed for $1+1$D transitions in the SymTFT setting to $2+1$D anyon condensation transitions. The construction applies when the anyons participating in the condensate have integer quantum dimensions. 
The second gives a criterion for identifying which point and loop excitations become critical at a continuous transition. The key step is to determine which anyon loops remain condensed throughout. All other loops become critical, while a point excitation becomes critical if it braids trivially with every condensed loop. We illustrate both proposals in several abelian and non-abelian examples. 
\end{abstract}

\maketitle


\section{Introduction}

Topologically ordered phases in $2+1$ dimensions are characterized by their anyonic excitations and how they fuse and braid. A transition between them can occur when a collection of bosonic anyons condenses. 
Their condensation reorganizes the remaining anyons through splitting, identification, and confinement, producing a new topological order. Mathematically, the condensing anyons and their fusion channels are encoded by a condensable algebra \cite{Bais2009condensate,Kong2014anyon,burnell2018anyon}. 

These transitions fall outside the conventional Ginzburg-Landau framework and are generally difficult to analyze. 
Exceptions are anyon condensation transitions that are related by gauging to ordinary group symmetry breaking ones. Microscopic Hamiltonians interpolating between topological orders related by anyon condensation have been constructed \cite{Christian2023lattice,Lin2023,Zhao2024nonabelian}, but they are exactly solvable only in the two limits. 
Numerical studies of such perturbed string-net models remain challenging \cite{Schulz2013fibonacci, Schotte2019tensor}, although Monte Carlo studies of quantum double models have made progress in this direction \cite{Flores-Calderon2026condensation}.

The anyon condensation picture has also been generalized to anyon proliferation, in which the participating quasiparticles need not be mutually bosonic \cite{Cheng2026abelianproliferation, Cheng2026peeling}.
A basic example is the simultaneous proliferation of the mutually nonlocal $e$ and $m$ excitations of the toric code, which has recently been revisited from a field-theoretic perspective \cite{Ji2026selfdual,Dumitrescu2026selfdual}.
A non-abelian counterpart was subsequently proposed for a self-dual transition in $D(S_3)$ \cite{Lu2026selfdual}. Systematic frameworks encompassing both condensation and proliferation have also been developed using the Symmetry Topological Field Theory formalism and Chern-Simons-matter theories \cite{Schafer-Nameki2026proliferation,Cheng2026proliferation, Garcia-Sepulveda:2026fud}.

Motivated by these developments, we revisit anyon condensation transitions and ask what can be learned about their critical behavior from the condensable algebra and the parent topological order. 

First, we seek a field theory description constructed directly from the condensable algebra. The starting point is the SymTFT-based procedure developed in our previous work on $1+1$D transitions \cite{Ravindran2026landau}. This approach requires the anyons in the condensable algebra to have integer quantum dimensions. We present the necessary modifications in Section~\ref{sec:GL_A} and illustrate the construction for the doubled Ising and $D(S_3)$ topological orders in Sections~\ref{sec:DIs} and~\ref{sec:DS3}.

Second, assuming the transition is continuous, we give a set of criteria on which excitations become critical. In $2+1$ dimensions, topological excitations, or defects, may be line- or pointlike. For a topologically ordered phase, closed anyon strings can be regarded as trivial, condensed loop excitations, while open anyon strings create nontrivial, gapped point excitations at their endpoints. 
At the transition, a loop is critical if the energy splitting produced by inserting it along a noncontractible cycle of the torus vanishes algebraically with system size; a point excitation is critical if its gap does the same. 
Our criterion about which loop and point excitations are critical at an anyon condensation transition is inspired by the SymTFT criterion for gapless excitations at $1+1$D transitions found in Ref.~\cite{Bhardwaj2025hasse}. After recasting the $1+1$D criterion in the language of topological defects in Section~\ref{sec:criterion1d}, we formulate our $2+1$D criterion in Section~\ref{sec:criterion2d} and give examples in Section~\ref{sec:criterionexamples}. We conclude in Section~\ref{sec:discussion} by summarizing our findings and outlining open questions towards a fusion 2-categorical description of point and line defects in $2+1$D critical systems.

\section{Ginzburg--Landau theory for anyon condensation transitions}
\label{sec:GL}

In Ref.~\cite{Ravindran2026landau}, we developed a generalized Ginzburg-Landau procedure to write down a field theory description of non-invertible symmetry breaking transitions in $1+1$D. Our discussion was based on the Symmetry Topological Field Theory (SymTFT) formalism\cite{Kong2015,Ji2020,Kong2020,Bhardwaj2020,Pulmann2021,Gaiotto2021,Lichtman2021,kong2020mathematical,Kong2020classification,Apruzzi2023,Chatterjee2023symmetry,Moradi2023,Freed2023topological,Lin2023,Kong2018,Kong2021,Kong2022one,Kong2022categories,kong2024categories,Xu2024,Bhardwaj2025} for $1+1$D systems with non-invertible symmetries and made extensive use of the topological data in the SymTFT formalism. 
The data includes the braided fusion category $\cC$ describing the $2+1$D topological order in the SymTFT bulk and the Lagrangian algebra $\cA$ describing the gapped boundary of the topological bulk. The same set of data describes the transition from the $2+1$D topological order of $\cC$ to the trivial phase through the condensation of $\cA$. Therefore, the procedure we developed in Ref.~\cite{Ravindran2026landau} can also be applied, if properly modified, to write down the field theory description of $2+1$D transitions between topological phases, driven by anyon condensation. This includes cases where the condensable algebra $\cA$ is not Lagrangian and the topological order after the transition is nontrivial. In sec.~\ref{sec:GL_A}, we explain basic steps of the construction where the anyons in $\cA$ play the role of order parameters of the transition. In sec.~\ref{sec:DIs} and sec.~\ref{sec:DS3}, we apply the procedure to two examples, one is the transition out of the Doubled Ising topological order and the other is the transition out of the $S_3$ Quantum Double topological order. 

\subsection{Construction from a condensable algebra}
\label{sec:GL_A}

Consider the phase transition from a $2+1$D topological phase driven by anyon condensation. The anyons in the bulk form a modular braided fusion category $\cC$ and are characterized by their fusion rule $a \times b = \sum_c N^{ab}_c c$, $N^{ab}_c \in \mathbb{Z}$ and exchange / braiding statistics. A subset of the anyons can form a condensable algebra $\cA = \sum_{a} n_a a$, $n_a \in \mathbb{Z}$ if they can consistently condense together. The procedure we describe below applies when the anyons in $\cA$ have integer quantum dimensions $d_a$ in the bulk. 

When the topological order is a gauge theory of group $G$ and the anyons condensed are (a subset of) the gauge charges, the transition can be `ungauged' into a normal symmetry breaking transition of symmetry $G$ where symmetry charges fluctuate and attain a nonzero expectation value in the symmetry-broken phase. The critical theory describing the topological phase transition is more or less the same as that of the $G$ symmetry breaking transition described by the usual Ginzburg-Landau field theory. They differ in terms of which symmetry sectors are included in the Hilbert space, but otherwise, the dynamical aspects are the same.

Generalizing this idea to other types of anyon condensation transitions, we propose that the condensing anyons $a \in \cA$ are the fluctuating order parameters of the transition. A Ginzburg-Landau field theory can be written for the transition if we can properly assign fields to the degrees of freedom associated with the condensing anyons and write down allowed Lagrangian terms using these degrees of freedom. Following the discussion in Ref.~\cite{Ravindran2026landau}, we propose a procedure that works when the quantum dimensions of the condensing anyons are integers.

\begin{itemize}
\item If an anyon $a \in \cA$ is not its own anti-particle (two $a$'s cannot fuse into identity), we assign complex fields to each of the internal dimensions of $a$. 

\item If $a \in \cA$ is its own anti-particle and two $a$'s fuse into identity in a way that is anti-symmetric under the exchange of the two $a$'s, we need to assign complex fields to the internal dimensions of $a$ but the fields are constrained.

\item If $a \in \cA$ is its own anti-particle and two $a$'s fuse into identity in a way that is symmetric under the exchange of the two $a$'s, we can assign real fields to the internal dimensions of $a$. The basis of $a$ to which real fields are assigned can be determined from the requirement that independent Lagrangian terms are real. 

\item Invariant Lagrangian terms of the order parameters correspond to fusion trees of anyons in $\cA$ that fuse into the trivial anyon $1$ (Fig.~\ref{fig:FusionTree}). Only fusion trees with all the branches in $\cA$ are allowed. The restriction of having all branches in $\cA$ is motivated by an RG picture in which the allowed fusion trees remain within $\cA$ at each coarse-graining step.

\begin{figure}[th]
\begin{center}
\begin{tikzpicture}[scale=0.7]
\begin{scope}
    \draw[dashed] (0,0) node[above right] {\large $1$} -- (0,1);
    \draw[] (0,1) -- (-1.,1.5);
    \draw[midarrow=0.8] (-1.,1.5) -- (-1.,2.5);
    \draw[] (0,1) -- (1.,1.5);
    \draw[midarrow=0.8] (1.,1.5) -- (1.,2.5);
    \node[left] at (-1.,2) {\large $a$};
    \node[right] at (1.,2) {\large $\bar a$};
\end{scope}
\begin{scope}[shift={(5,0)}]
    \draw[dashed] (0,0) -- (0,0.5);
    \node[right] at (0,0.2) {\large $1$};
    \draw[] (0,0.5) -- (-1,0.8);
    \draw[] (-1,0.8) -- (-1,1.3) -- (-2,1.8);
    \draw[] (-1,1.3) -- (0,1.8);
    \draw[midarrow=0.8] (-2,1.8) -- (-2,2.5);
    \draw[midarrow=0.8] (0,1.8) -- (0,2.5);
    \draw[] (0,0.5) -- (1,0.8);
    \draw[midarrow=0.8] (1,0.8) -- (1,2.5);
    \node[left] at (-2,2.15) {\large $a$};
    \node[left] at (0,2.15) {\large $b$};
    \node[left] at (1,2.15) {\large $c$};
\end{scope}   
\end{tikzpicture}
\caption{(a) Order two and (b) order three fusion trees of the condensing anyons that fuse into the trivial anyon $1$.} 
\label{fig:FusionTree}
\end{center}
\end{figure}
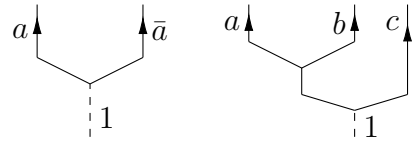

\item A field theory can be formulated by combining the invariant Lagrangian terms to describe a transition from the topological order $\cC$ to the topological order $\cC / \cA$ where the anyons in $\cA$ are condensed.

\item Different components of an order parameter may not be local with respect to each other. The mutual non-locality should be captured by topological terms in the Lagrangian. But for anyon condensation transitions, only a mutually-local subset eventually drives the transition while the other dimensions remain un-condensed.

\end{itemize}

In the next two sections, we illustrate how the procedure works with two examples: one is transition out of the Doubled Ising topological order (section~\ref{sec:DIs}), the other is the transition out of the Quantum Double of $S_3$ topological order (section~\ref{sec:DS3}).

\subsection{Example: Doubled Ising} 
\label{sec:DIs}

The Doubled Ising topological order consists of a time reversal pair of chiral topological orders. One chiral component contains anyons $\{1, \psi,\sigma\}$ which satisfy the fusion rule
\[
\psi\times \psi = 1, \ \psi\times \sigma = \sigma, \ \sigma\times \sigma = 1+\psi
\]
Its time reversal copy contains anyons $\{1, \bar{\psi},\bar{\sigma}\}$ which satisfy the same fusion rule but complex conjugated exchange and braiding statistics. The doubled anyons $\psi\bar{\psi}$ and $\sigma\bar{\sigma}$ are hence bosonic excitations, which can potentially be condensed. There are two condensable algebras inside the Doubled Ising topological order
\[
\cA_1 = 1 + \psi\bar{\psi}, \ \cA_2 = 1 + \psi\bar{\psi} + \sigma\bar{\sigma}
\]
The condensation driven by $\cA_1$ results in the $\mathbb{Z}_2$ Toric Code topological order, while the condensation driven by $\cA_2$ results in the trivial phase. Let's see how to write a field theory description for these two transitions following the procedure listed above.

The transition driven by $\cA_1$ involves the condensation of only abelian anyons and is well known to be in the $2+1$D Ising universality class \cite{Burnell2012}. The procedure above reproduces this result. The condensing anyon $\psi\bar{\psi}$ is an abelian $\mathbb{Z}_2$ anyon and should be associated with a real field $\phi$. To fuse into identity, we need an even number of $\psi\bar{\psi}$, resulting in Lagrangian terms $\phi^2$, $\phi^4$, etc. The field theory for the transition hence takes the form
\begin{equation}
\mathcal{L} = K (\partial\phi)^2 + \alpha \phi^2 + \beta \phi^4 + ...
\label{eq:L_Is}
\end{equation}
which describes the $2+1$D transition in the Ising universality class. After the condensation of $\cA_1$, the $\sigma\bar{\sigma}$ anyon splits into two abelian parts $e$ and $m$ which fuse into the fermion $\psi$. The topological order becomes that of the $\mathbb{Z}_2$ Toric Code.

The transition driven by $\cA_2$ is more complicated since it also involves the condensation of a non-abelian anyon $\sigma\bar{\sigma}$. $\sigma\bar{\sigma}$ has quantum dimension $2$ and is its own anti-particle. Therefore, we should assign two real fields to $\sigma\bar{\sigma}$, but we need to decide to which basis. Anticipating the splitting of $\sigma\bar{\sigma}$ into the $e$ and $m$ anyons of Toric Code, we label the two dimensions of $\sigma\bar{\sigma}$ as $\ket{e}$ and $\ket{m}$. $\ket{e}$ and $\ket{m}$ are exchanged when braided with anyon $\sigma$ or $\bar{\sigma}$. Both acquire a phase factor of $-1$ when braided with anyon $\psi$ or $\bar{\psi}$. Because of this, we can see that when two $\sigma\bar{\sigma}$'s fuse into the identity, they are entangled in the symmetric state $\ket{e}\ket{e} + \ket{m}\ket{m}$. If we assign field $\phi_e$ to $\ket{e}$ and $\phi_m$ to $\ket{m}$, the corresponding Lagrangian term takes the form $\phi_e^2+\phi_m^2$. For this term to be real, we see that both $\phi_e$ and $\phi_m$ are real.

Together with anyon $\psi\bar{\psi}$, there are two order-two fusion trees into identity
\[
\begin{tikzpicture}[baseline=-0.5ex, scale=0.6]
\begin{scope}
    \draw[densely dashed] (0,0) node[above right] {$1$} -- (0,1);
    \draw[] (0,1) -- (-1.,1.5);
    \draw[] (-1.,1.5) -- (-1.,2.5);
    \draw[] (0,1) -- (1.,1.5);
    \draw[] (1.,1.5) -- (1.,2.5);
    \node[left] at (-1.,2) {$\psi \bar \psi$};
    \node[right] at (1.,2) {$\psi \bar \psi$};
\end{scope}
\begin{scope}[shift={(5,0)}]
    \draw[densely dashed] (0,0) node[above right] {$1$} -- (0,1);
    \draw[] (0,1) -- (-1.,1.5);
    \draw[] (-1.,1.5) -- (-1.,2.5);
    \draw[] (0,1) -- (1.,1.5);
    \draw[] (1.,1.5) -- (1.,2.5);
    \node[left] at (-1.,2) {$\sigma \bar \sigma$};
    \node[right] at (1.,2) {$\sigma \bar \sigma$};
\end{scope}
\end{tikzpicture}
\]
which correspond to quantum states
\[
\ket{\psi\bar{\psi}}\ket{\psi\bar{\psi}}, \ \ket{e}\ket{e} + \ket{m}\ket{m}
\]
These states lead to Lagrangian terms
\[
\phi^2, \ \phi_e^2 + \phi_m^2
\]
and related derivative terms like $(\partial \phi)^2$.

At third order, there is one fusion tree into identity
\[
\begin{tikzpicture}[baseline=-0.5ex, scale=0.6]
    \draw[densely dashed] (0,-0.2) node[above right] {$1$} -- (0,0.5);
    \draw[] (0,0.5) -- (-1,0.8);
    \draw[] (-1,0.8) -- (-1,1.3) -- (-2,1.8);
    \draw[] (-1,1.3) -- (0,1.8);
    \draw[] (-2,1.8) -- (-2,2.5);
    \draw[] (0,1.8) -- (0,2.5);
    \draw[] (0,0.5) -- (1,0.8);
    \draw[] (1,0.8) -- (1,2.5);
    \node[left] at (-2,2.15) {$\sigma \bar \sigma$};
    \node[left] at (0,2.15) {$\sigma \bar \sigma$};
    \node[right] at (1,2.15) {$\psi \bar \psi$};
\end{tikzpicture}
\]
It corresponds to the quantum state
\[
\left(\ket{e}\ket{e}-\ket{m}\ket{m}\right)\ket{\psi\bar{\psi}}
\]
and results in the Lagrangian term
\[
\left(\phi_e^2-\phi_m^2\right)\phi
\]

At fourth order, there is one fusion tree involving four $\psi\bar{\psi}$'s and one fusion tree involving two $\psi\bar{\psi}$'s and two $\sigma\bar{\sigma}$'s
\[
\begin{tikzpicture}[baseline=-0.5ex, scale=0.6]
    \begin{scope}
    \draw[densely dashed] (0,-0.2) node[above right] {$1$} -- (0,0.5) -- (-1,0.8) -- (-1,1.3);
    \draw[] (-1,1.3) -- (-1.8,1.8);
    \draw[] (-1,1.3) -- (-0.2,1.8);
    \draw[] (-1.8,1.8) -- (-1.8,2.5);
    \draw[] (-0.2,1.8) -- (-0.2,2.5);
    \draw[densely dashed] (0,0.5) -- (1,0.8) -- (1,1.3); 
    \draw (1,1.3) -- (1.8,1.8) -- (1.8,2.5);
    \draw[] (1,1.3) -- (0.2,1.8) -- (0.2,2.5);
    \node[left] at (-1.8,2.15) {$\psi \bar \psi$};
    \node[left] at (-0.2,2.15) {$\psi \bar \psi$};
    \node[right] at (0.2,2.15) {$\psi \bar \psi$};
    \node[right] at (1.8,2.15) {$\psi \bar \psi$};
    \node[right] at (-1,1.1) {$1$};
    \node[left] at (1,1.1) {$1$};
    \end{scope}
    \begin{scope}[shift={(6.5,0)}]
    \draw[densely dashed] (0,-0.2) node[above right] {$1$} -- (0,0.5) -- (-1,0.8) -- (-1,1.3);
    \draw[] (-1,1.3) -- (-1.8,1.8);
    \draw[] (-1,1.3) -- (-0.2,1.8);
    \draw[] (-1.8,1.8) -- (-1.8,2.5);
    \draw[] (-0.2,1.8) -- (-0.2,2.5);
    \draw[densely dashed] (0,0.5) -- (1,0.8) -- (1,1.3); 
    \draw (1,1.3) -- (1.8,1.8) -- (1.8,2.5);
    \draw[] (1,1.3) -- (0.2,1.8) -- (0.2,2.5);
    \node[left] at (-1.8,2.15) {$\psi \bar \psi$};
    \node[left] at (-0.2,2.15) {$\psi \bar \psi$};
    \node[right] at (0.2,2.15) {$\sigma \bar \sigma$};
    \node[right] at (1.8,2.15) {$\sigma \bar \sigma$};
    \node[right] at (-1,1.1) {$1$};
    \node[left] at (1,1.1) {$1$};
    \end{scope}
\end{tikzpicture}
\]
They correspond respectively to quantum states
\[
\begin{array}{l}
\ket{\psi\bar{\psi}}\ket{\psi\bar{\psi}}\ket{\psi\bar{\psi}}\ket{\psi\bar{\psi}} \\
\ket{\psi\bar{\psi}}\ket{\psi\bar{\psi}}\left(\ket{e}\ket{e} + \ket{m}\ket{m}\right)
\end{array}
\]
and Lagrangian terms
\[
\phi^4, \ \phi^2\left(\phi_e^2+\phi_m^2\right)
\]

With four $\sigma\bar{\sigma}$'s, there are more fourth order fusion trees
\[
\begin{tikzpicture}[baseline=-0.5ex, scale=0.6]
    \begin{scope}
    \draw[densely dashed] (0,-0.2) node[above right] {$1$} -- (0,0.5) -- (-1,0.8) -- (-1,1.3);
    \draw[] (-1,1.3) -- (-1.8,1.8);
    \draw[] (-1,1.3) -- (-0.2,1.8);
    \draw[] (-1.8,1.8) -- (-1.8,2.5);
    \draw[] (-0.2,1.8) -- (-0.2,2.5);
    \draw[densely dashed] (0,0.5) -- (1,0.8) -- (1,1.3); 
    \draw (1,1.3) -- (1.8,1.8) -- (1.8,2.5);
    \draw[] (1,1.3) -- (0.2,1.8) -- (0.2,2.5);
    \node[left] at (-1.8,2.15) {$\sigma \bar \sigma$};
    \node[left] at (-0.2,2.15) {$\sigma \bar \sigma$};
    \node[right] at (0.2,2.15) {$\sigma \bar \sigma$};
    \node[right] at (1.8,2.15) {$\sigma \bar \sigma$};
    \node[right] at (-1,1.1) {$1$};
    \node[left] at (1,1.1) {$1$};
    \end{scope}
    \begin{scope}[shift={(6.5,0)}]
    \draw[densely dashed] (0,-0.2) node[above right] {$1$} -- (0,0.5);
    \draw (0,0.5) -- (-1,0.8) -- (-1,1.3) -- (-1.8,1.8);
    \draw[] (-1,1.3) -- (-0.2,1.8);
    \draw[] (-1.8,1.8) -- (-1.8,2.5);
    \draw[] (-0.2,1.8) -- (-0.2,2.5);
    \draw (0,0.5) -- (1,0.8) -- (1,1.3); 
    \draw (1,1.3) -- (1.8,1.8) -- (1.8,2.5);
    \draw[] (1,1.3) -- (0.2,1.8) -- (0.2,2.5);
    \node[left] at (-1.8,2.15) {$\sigma \bar \sigma$};
    \node[left] at (-0.2,2.15) {$\sigma \bar \sigma$};
    \node[right] at (0.2,2.15) {$\sigma \bar \sigma$};
    \node[right] at (1.8,2.15) {$\sigma \bar \sigma$};
    \node[left] at (-1,1.) {$\psi \bar \psi$};
    \node[right] at (1,1.) {$\psi \bar \psi$};
    \end{scope}
\end{tikzpicture}
\]
While each fusion tree satisfies the condition that all their branches are in $\cA_2$, this condition is not satisfied when the fusion order is changed using F moves. To make sure that the fusion tree contains only branches in $\cA_2$ for any fusion order, we need to sum over the two fusion trees above and obtain one fourth order fusion tree for four $\sigma\bar{\sigma}$'s.
\[
\begin{tikzpicture}[baseline=-0.5ex, scale=0.6]
    \begin{scope}
    \draw[densely dashed] (0,-0.2) node[above right] {$1$} -- (0,0.5) -- (-1,0.8) -- (-1,1.3);
    \draw[] (-1,1.3) -- (-1.8,1.8);
    \draw[] (-1,1.3) -- (-0.2,1.8);
    \draw[] (-1.8,1.8) -- (-1.8,2.5);
    \draw[] (-0.2,1.8) -- (-0.2,2.5);
    \draw[densely dashed] (0,0.5) -- (1,0.8) -- (1,1.3); 
    \draw (1,1.3) -- (1.8,1.8) -- (1.8,2.5);
    \draw[] (1,1.3) -- (0.2,1.8) -- (0.2,2.5);
    \node[left] at (-1.8,2.15) {$\sigma \bar \sigma$};
    \node[left] at (-0.2,2.15) {$\sigma \bar \sigma$};
    \node[right] at (0.2,2.15) {$\sigma \bar \sigma$};
    \node[right] at (1.8,2.15) {$\sigma \bar \sigma$};
    \node[right] at (-1,1.1) {$1$};
    \node[left] at (1,1.1) {$1$};
    \node[] at (3.3,1.) {$+$};
    \end{scope}
    \begin{scope}[shift={(6.5,0)}]
    \draw[densely dashed] (0,-0.2) node[above right] {$1$} -- (0,0.5);
    \draw (0,0.5) -- (-1,0.8) -- (-1,1.3) -- (-1.8,1.8);
    \draw[] (-1,1.3) -- (-0.2,1.8);
    \draw[] (-1.8,1.8) -- (-1.8,2.5);
    \draw[] (-0.2,1.8) -- (-0.2,2.5);
    \draw (0,0.5) -- (1,0.8) -- (1,1.3); 
    \draw (1,1.3) -- (1.8,1.8) -- (1.8,2.5);
    \draw[] (1,1.3) -- (0.2,1.8) -- (0.2,2.5);
    \node[left] at (-1.8,2.15) {$\sigma \bar \sigma$};
    \node[left] at (-0.2,2.15) {$\sigma \bar \sigma$};
    \node[right] at (0.2,2.15) {$\sigma \bar \sigma$};
    \node[right] at (1.8,2.15) {$\sigma \bar \sigma$};
    \node[left] at (-1,1.) {$\psi \bar \psi$};
    \node[right] at (1,1.) {$\psi \bar \psi$};
    \end{scope}
\end{tikzpicture}
\]
The summed fusion tree corresponds to the quantum state
\[
\ket{e}\ket{e}\ket{e}\ket{e}+\ket{m}\ket{m}\ket{m}\ket{m}
\]
The reduced density matrix of any two $\sigma\bar{\sigma}$'s is supported on dimensions $\ket{e}\ket{e}+\ket{m}\ket{m}$ and $\ket{e}\ket{e}-\ket{m}\ket{m}$, which upon fusion becomes $1$ and $\psi\bar{\psi}$, confirming that the fusion tree does not contain branches outside of $\cA_2$ no matter what the fusion order is. 

The Lagrangian term corresponding to the summed fusion tree is hence
\[
\phi_e^4+\phi_m^4
\]

Combining all the terms, we arrive at the field theory
\[
\begin{array}{lll}
\mathcal{L} & = &  K_1 (\partial\phi)^2 + \alpha_1 \phi^2 + \beta_1 \phi^4  \\
& & + \cL_{\text{top}} + K_2 [(D_1\phi_e)^2 + (D_2\phi_m)^2] + \alpha_2 \left(\phi_e^2 + \phi_m^2\right) \\
& & + \beta_2 \left(\phi_e^2-\phi_m^2\right)\phi + \gamma \phi^2\left(\phi_e^2 + \phi_m^2\right)  \\
& & + \delta \left(\phi_e^4 + \phi_m^4\right) + ...
\end{array}
\]
which has a similar form to the abelian double Chern-Simons theory in \cite{Ji2026selfdual}. The first row contains terms involving only $\phi$ and is the usual Lagrangian of the Ising transition. The second row and the fourth row contain terms involving $\phi_e$ and $\phi_m$. The nontrivial feature of $\phi_e$ and $\phi_m$ is that they are mutually non-local. Their mutual non-locality is captured by the topological term $\cL_{\text{top}}$
\[
\cL_{\text{top}} = i\frac{2}{2\pi} a_1 \wedge da_2 + \sum_{i=1}^2\frac{1}{4e^2} \left(\partial \times a_i\right)^2
\]
$\phi_e$ and $\phi_m$ are coupled to the gauge fields $a_1$ and $a_2$ respectively through the covariant derivatives $D_1$ and $D_2$. The mutual Chern-Simons term $i\frac{2}{2\pi} a_1 \wedge da_2$ attaches a $\pi$ flux of $a_1$ to a unit charge of $\phi_m$ and vice versa, thereby imposing a $-1$ braiding statistics between $\phi_e$ and $\phi_m$.
This topological term is important in the field theory description when $\phi_e$ and $\phi_m$ are equally important in driving the transition. But since $\phi_e$ and $\phi_m$ do not commute, they cannot condense together. If the transition is an anyon condensation transition, then it is driven by the condensation either of $\phi_e$ or $\phi_m$, but not both. This happens when the symmetry exchanging $\phi_e$ and $\phi_m$ is broken, which can be induced through the coupling term in the third row to $\phi$.

Suppose that the transition splits into two steps. In the first step, $\alpha_1$ becomes negative and $\phi$ attains a nonzero expectation value. When this happens, the first term in the third row (the $\beta_2$ term) becomes effectively quadratic terms for $\phi_e$ and $\phi_m$ but with different coefficients. Then in the second step of the transition, $\alpha_2$ becomes negative. The coefficients of $\phi_e^2$ and $\phi_m^2$ vary as a result, but one of them becomes negative first. That is, one of $\phi_e$ or $\phi_m$ drives the transition while the other one remains gapped. In this scenario, the mutual non-locality between $\phi_e$ and $\phi_m$ is not that important and the transition splits into two Ising type transitions, which is well understood.

\subsection{Example: Quantum Double \texorpdfstring{$D(S_3)$}{D(S3)}}
\label{sec:DS3}

$S_3$ is the non-abelian symmetry group of a triangle. It is generated by a three-fold rotation operation $f$ ($f^3=1$) and a reflection operation $d_1$ ($d_1^2=1$). $f$ and $d_1$ satisfy the relation $(fd_1)^2 = 1$. We will use the notation $\bar{f}=f^2$, $d_2=fd_1$, and $d_3=\bar{f}d_1$. There are eight types of anyons in the Quantum Double topological order of $S_3$ labeled $A$ through $H$\cite{Kitaev2003,Beigi2011}. $A$, $B$, $C$ are gauge charges corresponding to the identity irrep, the sign irrep and the two-dimensional irrep of $S_3$. $D$ and $F$ are gauge fluxes corresponding to the reflection and the rotation conjugacy classes of $S_3$. $E$, $G$, $H$ are the dyons -- the charge-flux composites. The gauge charges $A$, $B$, $C$ and gauge fluxes $D$, $F$ are bosonic and can potentially be condensed to drive an anyon condensation transition. 

There are seven condensable algebras in the Quantum Double of $S_3$. 
\[
\begin{array}{l}
\cA_1 = A+B, \ \cA_2 = A+C,\ \cA_3 = A+B+2C \\
\cA_4 = A+F,\ \cA_5 = A+B+2F \\
\cA_6 = A+D+C, \ \cA_7 = A+D+F
\end{array}
\]
Due to the duality between the $C$ and $F$ anyon, $\cA_2$ is dual to $\cA_4$, $\cA_3$ is dual to $\cA_5$, and $\cA_6$ is dual to $\cA_7$. In the following, we will write down the field theory description of the transitions driven by the condensation of $\cA_1$, $\cA_2$, $\cA_3$, and $\cA_6$.

The condensation driven by $\cA_1 = A+B$ involves only one nontrivial anyon $B$ which is abelian and of order 2, $B\times B=A$. To write down the transition field theory, we assign a real field $\phi_b$ to $B$ and allow all even order terms $\phi_b^2$, $\phi_b^4$, .... The resulting field theory is the Ising field theory in Eq.~\ref{eq:L_Is}, with $\phi$ replaced by $\phi_b$.
\begin{equation}
\mathcal{L}_1 = K (\partial\phi_b)^2 + \alpha \phi_b^2 + \beta \phi_b^4 + ...
\end{equation}
After the condensation, the topological order is reduced to that of a $\mathbb{Z}_3$ Toric Code, which is the gauge theory of the abelian group $\mathbb{Z}_3$. 

The condensation driven by $\cA_2 = A+C$ also involves only one nontrivial anyon, but it is more complicated since $C$ is non-abelian with quantum dimension 2. Since $C$ is its own anti-particle, we can assign two real fields to capture its fluctuation. Label the two internal dimensions of $C$ as $\ket{c}$ and $\ket{\bar{c}}$, which correspond to the charge $1$ and $-1$ dimensions under the $\mathbb{Z}_3$ rotation symmetry within the 2-dimensional irrep. To figure out how the two real fields should be assigned, notice that when two $C$'s fuse into $A$, the singlet state takes the form 
\[
\ket{c}\ket{\bar{c}} + \ket{\bar{c}}\ket{c}
\]
For the corresponding Lagrangian term to be real, we assign a complex field $\psi_c$ to $\ket{c}$ and its complex conjugate $\bar{\psi}_c$ to $\ket{\bar{c}}$. The quadratic Lagrangian term is then
\[
|\psi_c|^2
\]
At third order, there is one fusion tree of three $C$'s
\[
\begin{tikzpicture}[baseline=-0.5ex, scale=0.6]
    \draw[densely dashed] (0,-0.2) node[above right] {$A$} -- (0,0.5);
    \draw[] (0,0.5) -- (-1,0.8);
    \draw[] (-1,0.8) -- (-1,1.3) -- (-2,1.8);
    \draw[] (-1,1.3) -- (0,1.8);
    \draw[] (-2,1.8) -- (-2,2.5);
    \draw[] (0,1.8) -- (0,2.5);
    \draw[] (0,0.5) -- (1,0.8);
    \draw[] (1,0.8) -- (1,2.5);
    \node[left] at (-2,2.15) {$C$};
    \node[left] at (0,2.15) {$C$};
    \node[right] at (1,2.15) {$C$};
    \node[right] at (-1,1.1) {$C$};
\end{tikzpicture}
\]
which corresponds to quantum state
\[
\ket{c}\ket{c}\ket{c} + \ket{\bar{c}}\ket{\bar{c}}\ket{\bar{c}}
\]
and Lagrangian term
\[
\psi_c^3 + \bar{\psi}_c^3
\]
At fourth order, we form a linear superposition of two fusion trees to make sure the fusion branches are always within $\cA_2$.
\[
\begin{tikzpicture}[baseline=-0.5ex, scale=0.6]
    \begin{scope}
    \draw[densely dashed] (0,-0.2) node[above right] {$A$} -- (0,0.5) -- (-1,0.8) -- (-1,1.3);
    \draw[] (-1,1.3) -- (-1.8,1.8);
    \draw[] (-1,1.3) -- (-0.2,1.8);
    \draw[] (-1.8,1.8) -- (-1.8,2.5);
    \draw[] (-0.2,1.8) -- (-0.2,2.5);
    \draw[densely dashed] (0,0.5) -- (1,0.8) -- (1,1.3); 
    \draw (1,1.3) -- (1.8,1.8) -- (1.8,2.5);
    \draw[] (1,1.3) -- (0.2,1.8) -- (0.2,2.5);
    \node[left] at (-1.8,2.15) {$C$};
    \node[left] at (-0.2,2.15) {$C$};
    \node[right] at (0.2,2.15) {$C$};
    \node[right] at (1.8,2.15) {$C$};
    \node[right] at (-1,1.1) {$A$};
    \node[left] at (1,1.1) {$A$};
    \node[] at (3.3,1.) {$+$};
    \end{scope}
    \begin{scope}[shift={(6.5,0)}]
    \draw[densely dashed] (0,-0.2) node[above right] {$A$} -- (0,0.5);
    \draw (0,0.5) -- (-1,0.8) -- (-1,1.3) -- (-1.8,1.8);
    \draw[] (-1,1.3) -- (-0.2,1.8);
    \draw[] (-1.8,1.8) -- (-1.8,2.5);
    \draw[] (-0.2,1.8) -- (-0.2,2.5);
    \draw (0,0.5) -- (1,0.8) -- (1,1.3); 
    \draw (1,1.3) -- (1.8,1.8) -- (1.8,2.5);
    \draw[] (1,1.3) -- (0.2,1.8) -- (0.2,2.5);
    \node[left] at (-1.8,2.15) {$C$};
    \node[left] at (-0.2,2.15) {$C$};
    \node[right] at (0.2,2.15) {$C$};
    \node[right] at (1.8,2.15) {$C$};
    \node[right] at (-1,1.1) {$C$};
    \node[left] at (1,1.1) {$C$};
    \end{scope}
\end{tikzpicture}
\]
which leads to quantum state
\[
\ket{c}\ket{c}\ket{\bar{c}}\ket{\bar{c}} + \text{permutation}
\]
and Lagrangian term
\[
|\psi_c|^4
\]
Putting all the terms together, we arrive at 
\[
\begin{array}{lll}
\mathcal{L}_2 & = &  K |\partial \psi_c|^2 + \alpha |\psi_c|^2 \\
& &  + \beta (\psi_c^3 + \bar{\psi}_c^3) + \gamma |\psi_c|^4 + ...
\end{array}
\]
which describes a $\mathbb{Z}_3$ symmetry-breaking transition in $2+1$D. After the transition, the topological order is reduced to that of a $\mathbb{Z}_2$ Toric Code, which is the gauge theory of the abelian group $\mathbb{Z}_2$.

Unlike $\cA_1$ and $\cA_2$, the algebra $\cA_3 = A+B+2C$ is Lagrangian, meaning that the condensation of $\cA_3$ drives the transition to a trivial topological phase. Both $B$ and $C$ are order parameters of this transition. The Lagrangian includes terms with $B$ only and those with $C$ only, as well as those that mix $B$ and $C$. In particular, a third order mixed fusion tree involves one $B$ and two $C$'s
\[
\begin{tikzpicture}[baseline=-0.5ex, scale=0.6]
    \draw[densely dashed] (0,-0.2) node[above right] {$A$} -- (0,0.5);
    \draw[] (0,0.5) -- (-1,0.8);
    \draw[] (-1,0.8) -- (-1,1.3) -- (-2,1.8);
    \draw[] (-1,1.3) -- (0,1.8);
    \draw[] (-2,1.8) -- (-2,2.5);
    \draw[] (0,1.8) -- (0,2.5);
    \draw[] (0,0.5) -- (1,0.8);
    \draw[] (1,0.8) -- (1,2.5);
    \node[left] at (-2,2.15) {$C$};
    \node[left] at (0,2.15) {$C$};
    \node[right] at (1,2.15) {$B$};
    \node[right] at (-1,1.1) {$B$};
\end{tikzpicture}
\]
The fusion tree corresponds to the quantum state
\[
\left(\ket{c}\ket{\bar{c}}-\ket{\bar{c}}\ket{c}\right)\ket{b}
\]
If we replace the two $C$'s with the same fields, the resulting Lagrangian term would be zero. However, if we associate each $C$ with a different field, e.g. by allowing spatial variation from one $C$ to the other, we obtain the term 
\[
\phi_b \left(\psi_c\partial_i\bar{\psi}_c-\bar{\psi}_c\partial_i\psi_c\right), \ i = x,y 
\]
Putting the terms together, we arrive at the Lagrangian
\[
\cL_3 = \cL_1 + \cL_2 + \sum_{i=x,y}\lambda_i \phi_b \left(\psi_c\partial_i\bar{\psi}_c-\bar{\psi}_c\partial_i\psi_c\right) + ...
\]
With two order parameters, this transition can be split into two parts. First, the quadratic coefficient of $\phi_b$ goes negative, driving an Ising type transition by condensing $B$. Then the quadratic coefficient of $\psi_c$ goes negative, driving a $\mathbb{Z}_3$ symmetry-breaking type transition by condensing $C$. With a nonzero expectation value of $\phi_b$, the mixed third order term in $\cL_3$ becomes a chiral term in the $\mathbb{Z}_3$ symmetry-breaking transition. In $1+1$D, this chiral term can change the nature of the $\mathbb{Z}_3$ symmetry-breaking transition point\cite{Huse1982,Chepiga2019,Maceira2022}, although in $2+1$D, we expect a first order transition with or without the chiral term.

Finally, let's consider the transition driven by $\cA_6 = A+D+C$. The Lagrangian terms involving only $C$ are the same as those of $\cA_2$. We just need to find the Lagrangian terms that involve only $D$ and those that involve a mixture of $D$ and $C$. $D$ has quantum dimension 3 and is its own anti-particle. Therefore, we should assign three real fields to $D$. To decide which basis states should be associated with these real fields, we note that when two $D$'s fuse into $A$, they form a symmetric entangled state
\[
\ket{d_1}\ket{d_1}+\ket{d_2}\ket{d_2}+\ket{d_3}\ket{d_3}
\]
where $d_1$, $d_2$, $d_3$ are the three group elements in the reflection conjugacy class. To ensure that the corresponding Lagrangian term is real, we assign real fields $\phi_{d_1}$, $\phi_{d_2}$, $\phi_{d_3}$ to $\ket{d_1}$, $\ket{d_2}$, and $\ket{d_3}$ and obtain quadratic Lagrangian term
\[
\phi_{d_1}^2 + \phi_{d_2}^2 + \phi_{d_3}^2
\]\

At third order, there is a mixed fusion tree of two $D$'s and one $C$
\[
\begin{tikzpicture}[baseline=-0.5ex, scale=0.6]
    \draw[densely dashed] (0,-0.2) node[above right] {$A$} -- (0,0.5);
    \draw[] (0,0.5) -- (-1,0.8);
    \draw[] (-1,0.8) -- (-1,1.3) -- (-2,1.8);
    \draw[] (-1,1.3) -- (0,1.8);
    \draw[] (-2,1.8) -- (-2,2.5);
    \draw[] (0,1.8) -- (0,2.5);
    \draw[] (0,0.5) -- (1,0.8);
    \draw[] (1,0.8) -- (1,2.5);
    \node[left] at (-2,2.15) {$D$};
    \node[left] at (0,2.15) {$D$};
    \node[right] at (1,2.15) {$C$};
    \node[right] at (-1,1.1) {$C$};
\end{tikzpicture}
\]
which corresponds to quantum state
\[
\begin{array}{l}
\left(\ket{d_1}\ket{d_1}+\omega\ket{d_2}\ket{d_2}+\bar{\omega}\ket{d_3}\ket{d_3}\right)\ket{c}+ \\
\left(\ket{d_1}\ket{d_1}+\bar{\omega}\ket{d_2}\ket{d_2}+\omega\ket{d_3}\ket{d_3}\right)\ket{\bar{c}}
\end{array}
\]
and Lagrangian term
\[
\left(\phi_{d_1}^2+\omega\phi_{d_2}^2+\bar{\omega}\phi_{d_3}^2\right)\psi_c+ 
\left(\phi_{d_1}^2+\bar{\omega}\phi_{d_2}^2+\omega\phi_{d_3}^2\right)\bar{\psi}_c
\]

At fourth order, there is one fusion tree involving four $D$'s,
\[
\begin{tikzpicture}[baseline=-0.5ex, scale=0.6]
    \begin{scope}
    \draw[densely dashed] (0,-0.2) node[above right] {$A$} -- (0,0.5) -- (-1,0.8) -- (-1,1.3);
    \draw[] (-1,1.3) -- (-1.8,1.8);
    \draw[] (-1,1.3) -- (-0.2,1.8);
    \draw[] (-1.8,1.8) -- (-1.8,2.5);
    \draw[] (-0.2,1.8) -- (-0.2,2.5);
    \draw[densely dashed] (0,0.5) -- (1,0.8) -- (1,1.3); 
    \draw (1,1.3) -- (1.8,1.8) -- (1.8,2.5);
    \draw[] (1,1.3) -- (0.2,1.8) -- (0.2,2.5);
    \node[left] at (-1.8,2.15) {$D$};
    \node[left] at (-0.2,2.15) {$D$};
    \node[right] at (0.2,2.15) {$D$};
    \node[right] at (1.8,2.15) {$D$};
    \node[right] at (-1,1.1) {$A$};
    \node[left] at (1,1.1) {$A$};
    \node[] at (3.3,1.) {$+$};
    \end{scope}
    \begin{scope}[shift={(6.5,0)}]
    \draw[densely dashed] (0,-0.2) node[above right] {$A$} -- (0,0.5);
    \draw (0,0.5) -- (-1,0.8) -- (-1,1.3) -- (-1.8,1.8);
    \draw[] (-1,1.3) -- (-0.2,1.8);
    \draw[] (-1.8,1.8) -- (-1.8,2.5);
    \draw[] (-0.2,1.8) -- (-0.2,2.5);
    \draw (0,0.5) -- (1,0.8) -- (1,1.3); 
    \draw (1,1.3) -- (1.8,1.8) -- (1.8,2.5);
    \draw[] (1,1.3) -- (0.2,1.8) -- (0.2,2.5);
    \node[left] at (-1.8,2.15) {$D$};
    \node[left] at (-0.2,2.15) {$D$};
    \node[right] at (0.2,2.15) {$D$};
    \node[right] at (1.8,2.15) {$D$};
    \node[right] at (-1,1.1) {$C$};
    \node[left] at (1,1.1) {$C$};
    \end{scope}
\end{tikzpicture},
\]
and one involving two $D$'s and two $C$'s
\[
\begin{tikzpicture}[baseline=-0.5ex, scale=0.6]
    \begin{scope}
    \draw[densely dashed] (0,-0.2) node[above right] {$A$} -- (0,0.5) -- (-1,0.8) -- (-1,1.3);
    \draw[] (-1,1.3) -- (-1.8,1.8);
    \draw[] (-1,1.3) -- (-0.2,1.8);
    \draw[] (-1.8,1.8) -- (-1.8,2.5);
    \draw[] (-0.2,1.8) -- (-0.2,2.5);
    \draw[densely dashed] (0,0.5) -- (1,0.8) -- (1,1.3); 
    \draw (1,1.3) -- (1.8,1.8) -- (1.8,2.5);
    \draw[] (1,1.3) -- (0.2,1.8) -- (0.2,2.5);
    \node[left] at (-1.8,2.15) {$C$};
    \node[left] at (-0.2,2.15) {$C$};
    \node[right] at (0.2,2.15) {$D$};
    \node[right] at (1.8,2.15) {$D$};
    \node[right] at (-1,1.1) {$A$};
    \node[left] at (1,1.1) {$A$};
    \node[] at (3.3,1.) {$+$};
    \end{scope}
    \begin{scope}[shift={(6.5,0)}]
    \draw[densely dashed] (0,-0.2) node[above right] {$A$} -- (0,0.5);
    \draw (0,0.5) -- (-1,0.8) -- (-1,1.3) -- (-1.8,1.8);
    \draw[] (-1,1.3) -- (-0.2,1.8);
    \draw[] (-1.8,1.8) -- (-1.8,2.5);
    \draw[] (-0.2,1.8) -- (-0.2,2.5);
    \draw (0,0.5) -- (1,0.8) -- (1,1.3); 
    \draw (1,1.3) -- (1.8,1.8) -- (1.8,2.5);
    \draw[] (1,1.3) -- (0.2,1.8) -- (0.2,2.5);
    \node[left] at (-1.8,2.15) {$C$};
    \node[left] at (-0.2,2.15) {$C$};
    \node[right] at (0.2,2.15) {$D$};
    \node[right] at (1.8,2.15) {$D$};
    \node[right] at (-1,1.1) {$C$};
    \node[left] at (1,1.1) {$C$};
    \end{scope}
\end{tikzpicture}.
\]
Again the fusion trees are superposed so that all branches are within $\cA_6$ no matter what the fusion order is. The fusion trees correspond to quantum states
\[
\begin{array}{l}
\ket{d_1d_1d_1d_1}+\ket{d_2d_2d_2d_2}+\ket{d_3d_3d_3d_3},
\\
\left(\ket{d_1d_1}+\ket{d_2d_2}+\ket{d_3d_3}\right)\left(\ket{c\bar{c}}+\ket{\bar{c}c}\right) \\
+ \left(\ket{d_1d_1}+\omega\ket{d_2d_2}+\bar{\omega}\ket{d_3d_3}\right)\ket{\bar{c}\bar{c}} \\
+ \left(\ket{d_1d_1}+\bar{\omega}\ket{d_2d_2}+\omega\ket{d_3d_3}\right)\ket{cc}
\end{array}
\]
and the corresponding Lagrangian terms are
\[
\begin{array}{l}
\phi_{d_1}^4 + \phi_{d_2}^4 + \phi_{d_3}^4, \\
\left(\phi_{d_1}^2 + \phi_{d_2}^2 + \phi_{d_3}^2\right)|\psi_c|^2 \\
+ \left(\phi_{d_1}^2 + \omega\phi_{d_2}^2 + \bar{\omega}\phi_{d_3}^2\right)\bar{\psi_c}^2 \\
+ \left(\phi_{d_1}^2 + \bar{\omega}\phi_{d_2}^2 + \omega\phi_{d_3}^2\right)\psi_c^2
\end{array}
\]

Now if we put all the terms together, we obtain the field theory describing the transition driven by $\cA_6$
\[
\begin{array}{lll}
\cL_6 & = & K_1 |\partial \psi_c|^2 + \alpha_1 |\psi_c|^2 + \beta_1 (\psi_c^3 + \bar{\psi}_c^3) + \gamma_1 |\psi_c|^4 \\
& & + \cL_{\text{top}} + K_2 \sum_i \left(\partial \phi_{d_i}\right)^2 + \alpha_2 \sum_i \phi_{d_i}^2 + \beta_2 \sum_i \phi_{d_i}^4 \\
& & + \alpha_3 \left(\phi_{d_1}^2 + \phi_{d_2}^2 + \phi_{d_3}^2\right)|\psi_c|^2 \\
& & + \left(\phi_{d_1}^2 + \omega\phi_{d_2}^2 + \bar{\omega}\phi_{d_3}^2\right)\left(\beta_3 \psi_c + \alpha_3 \bar{\psi_c}^2\right) \\
& & + \left(\phi_{d_1}^2 + \bar{\omega}\phi_{d_2}^2 + \omega\phi_{d_3}^2\right)\left(\beta_3\bar{\psi}_c + \alpha_3\psi_c^2\right) \\
& & + ...
\end{array}
\]
Note that the $\ket{d_1}$, $\ket{d_2}$, $\ket{d_3}$ dimensions are not mutually local with respect to each other. Therefore, a topological term $\cL_{\text{top}}$ is needed to capture this. We will not try to write down the explicit form of the topological term. Since the $\ket{d_i}$ dimensions are not mutually local, only one of them can condense at a time while the other two remain gapped. Therefore, at the anyon condensation transition, the topological term is not important.

The transition driven by $\cA_6$ can split into two parts. First, the quadratic coefficient of $\psi_c$ goes negative and drives the condensation of $C$. This is a $\mathbb{Z}_3$ symmetry-breaking type of transition. After this transition, $\psi_c$ acquires a nonzero expectation value, and the mixed terms between $D$ and $C$ become effectively quadratic terms for $D$ but with different coefficients for the $d_i$'s. Therefore, when $D$ further condenses, one of the components, say $d_1$, goes through the transition first while $d_2$ and $d_3$ remain gapped. This part of the transition is of the Ising type.

\section{Critical excitations at \texorpdfstring{$2+1$D}{2+1D} anyon-condensation transitions}

Writing down a field theory can be a useful first step in understanding a transition, but field theory can be hard to analyze and does not always tell us directly about universal properties of the transition. An ideal situation is that we can directly extract universal features of a transition from the universal features of its neighboring gapped phases. This might be too much to ask, but some universal features of the critical point might be easier to identify than others. In this section, we propose a way to generalize the discussion in $1+1$D~\cite{Bhardwaj2025hasse, Chatterjee2023holographic} and show how properties of topological defects at the transition can be predicted from their behavior in the neighboring gapped phases. A crucial feature in our $2+1$D discussion is the 2-category structure of topological defects in $2+1$D. That is, topological defects in $2+1$D come at different levels. At the first level are $1+1$D defects, which correspond to $0$-form symmetries. At the second level are $0+1$D defects that live at the domain wall between $1+1$D defects. In particular, when the $1+1$D defects are condensed, the $0+1$D defects become truly isolated defects and correspond to $1$-form symmetries. This approach is inspired by the 2-category structure of topological excitations in $3+1$D topological orders\cite{Johnson-Freyd2022,Kong2020classification,Kong2014arxiv}, although in our discussion we do not use the SymTFT formalism where a $2+1$D system is realized in a sandwich structure with a $3+1$D topological order in the bulk.

\subsection{Criterion for critical excitations in \texorpdfstring{$1+1$D}{1+1D}}\label{sec:criterion1d}

Here we review a criterion for determining which topological defects of neighbouring gapped phases become gapless at a critical transition. The criterion was stated in the SymTFT framework in Refs.~\cite{Bhardwaj2025hasse, Chatterjee2023holographic}; we recast it in the language of low-entanglement excitations \cite{Ji2025bulk}, or topological defects. In $1+1$D, these two formulations are equivalent. We do not use the SymTFT framework for the review because in $2+1$D the SymTFT framework is limited, while the topological defect perspective will allow us to discuss anyon condensation transitions in more generality. 

The low-entanglement excitations \cite{Ji2025bulk}, or topological defects in gapped phases, are $d$-dimensional excitations on top of a $D$-dimensional gapped ground state, with $d < D$, that preserve its area law. Two such excitations are in the same equivalence class if they are connected by a $d$-dimensional finite-depth circuit composed of symmetry-preserving gates. For a $1+1$D Hamiltonian, all these excitations are $0+1$D point-like defects. Nontrivial defects include symmetry charges and symmetry fluxes. Symmetry charges are excitations created by local operators that transform nontrivially under the symmetry. Symmetry fluxes correspond to ground states of Hamiltonians modified by a symmetry twist; equivalently, they can be inserted by nonlocal string operators.
In the SymTFT, each topological defect type is labeled by an anyon in the $2+1$D bulk. A non-abelian anyon may give rise to a multiplet of defects that are mixed by the symmetry action. For a non-invertible symmetry, this action can also mix local and nonlocal operators.

A gapped phase is characterized by which topological defects are condensed and how they condense. For a multiplet of defects, only a particular component or subspace may condense. Several defects can condense together only if they are mutually local, meaning that the corresponding defect operations commute. Condensed symmetry charges give local order parameters and determine the ground state degeneracy, while condensed symmetry fluxes give string order parameters.

\begin{criterion}\label{criterion1d}
    Consider a direct continuous transition between two $1+1$D gapped phases, and let $K$ denote the set of topological defects condensed in both phases. Then:
    \begin{enumerate}
        \item Defects in $K$ remain condensed at the transition.
        \item Defects outside $K$ that are mutually local with every defect in $K$ become critical at the transition. Their finite-size excitation gaps scale as $L^{-z}$, in particular as $1/L$ for a 2D CFT.
        \item Defects outside $K$ that are not mutually local with $K$ stay gapped throughout the transition.
    \end{enumerate}
\end{criterion}

\begin{figure}[th]
\begin{center}
\includegraphics[width= \textwidth]{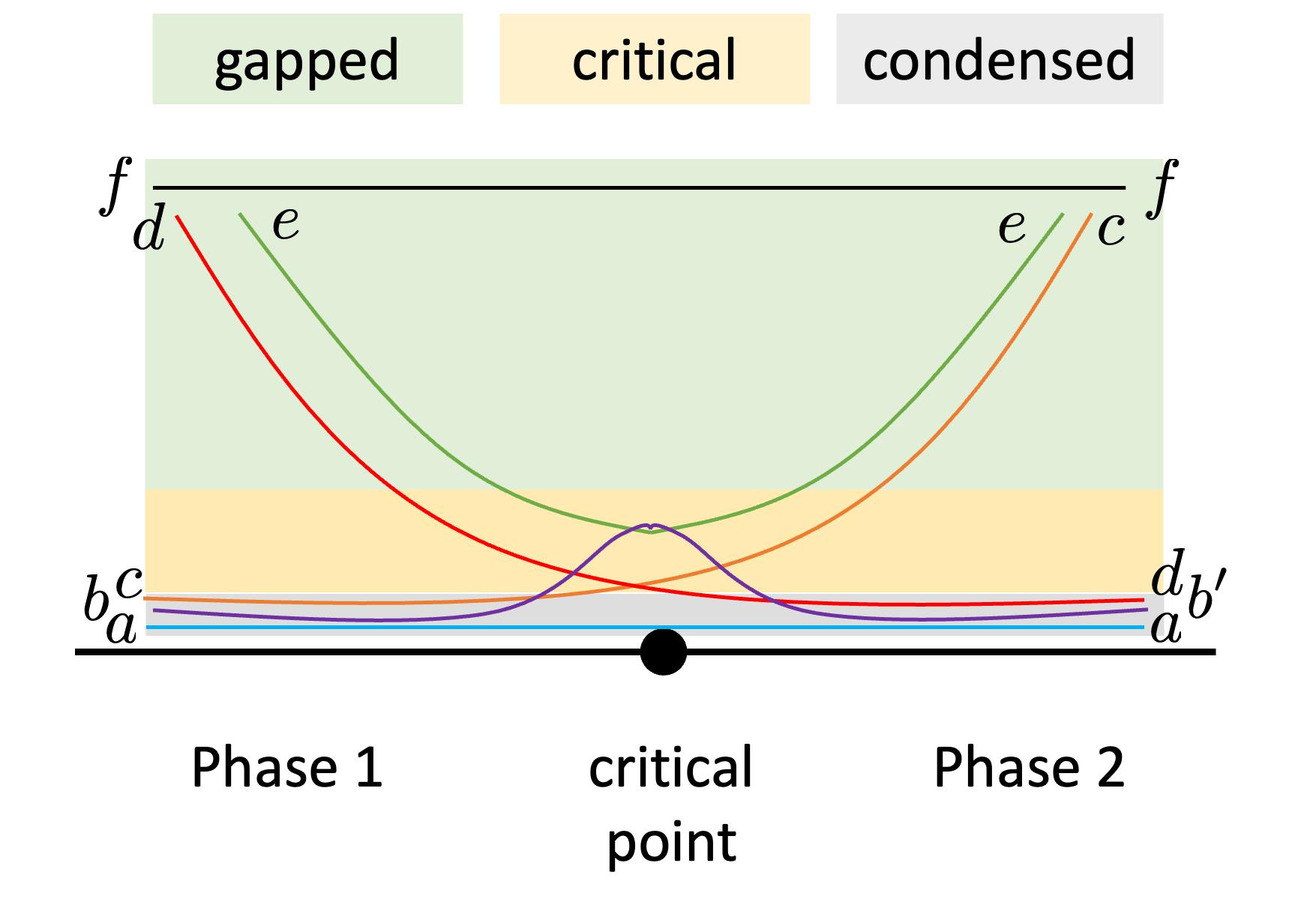}
\caption{Schematic behaviour of topological defects across a continuous transition. In the gapped phases on the two sides of the critical point, defects are either condensed or gapped. At the critical point, some defects remain condensed, some remain gapped, while others become critical.} 
\label{fig:defects}
\end{center}
\end{figure}

In Fig.~\ref{fig:defects}, we illustrate the behavior of different types of defects across a transition. Defects in the gray region are condensed. That is, their energy is exponentially small in the system size and their correlation length is infinite. Defects in the yellow region are critical: their energy scales as an inverse polynomial of system size and their correlation function also decays as a polynomial. Defects in the green region are gapped: their energy is a finite value above zero and they have a finite correlation length. Inside the gapped phases, defects are either condensed or gapped. At the continuous transition point between the gapped phases, some of the defects become critical, according to the criterion stated above.

For example, $a$ is a defect in $K$. It is condensed in both gapped phases and remains condensed at the critical point. $b$ is condensed in both gapped phases, but with different components in each phase. We indicate this difference with $b$ and $b'$. Because of the change in condensed component, $b$ is not in $K$. On the other hand, it is mutually local with any defects in $K$ because in each gapped phase, it is condensed together with $K$. Therefore, $b$ becomes critical at the transition. We will discuss an example of $S_3$ symmetry breaking transition below where a defect of this type appears. $c$ and $d$ are condensed in one of the phases but gapped in the other. Since they are condensed together with $K$ in one of the phases, they become critical at the transition. $e$ and $f$ are gapped in both phases, but their behavior at the transition is different. $f$ is non-local with respect to some objects in $K$ and therefore remains gapped at the transition. $e$ is mutually local with all the objects in $K$ and therefore becomes critical at the transition. 

In the SymTFT language, a direct continuous transition between two gapped phases with physical boundaries specified by Lagrangian algebras $\cA_1$ and $\cA_2$ is described by a non-maximal condensable algebra $\cK$ satisfying $\cK \subset \cA_1$ and $\cK \subset \cA_2$. The above criterion is stated as: the anyons contained in $\cK$ remain condensed at the critical point; any other bulk anyon with trivial braiding with every anyon in $\cK$ becomes gapless, whereas anyons with nontrivial braiding with $\cK$ remain gapped \cite{Bhardwaj2025hasse}.

As an example, consider the three-state quantum Potts model
\begin{equation}
H=-J\sum_j\left(\sigma_j^\dagger\sigma_{j+1}+\mathrm{h.c.}\right)-f\sum_j\left(\tau_j+\tau_j^\dagger\right),
\end{equation}
where $\sigma_j$ and $\tau_j$ are the three dimensional clock and shift operators that satisfy $\sigma_j\tau_j=\omega\tau_j\sigma_j$, $\omega=e^{2\pi i/3}$. It has an $S_3$ symmetry generated by the $\mathbb Z_3$ rotation 
\begin{equation}
U_f=\prod_j\tau_j^\dagger, \qquad U_f\sigma_j U_f^\dagger=\omega\sigma_j ,
\end{equation}
and charge conjugation
\begin{equation}
U_{d_1}=\prod_j s_j,\qquad s_j \, (\sigma_j,\tau_j) \, s_j^{-1} =(\sigma_j^\dagger,\tau_j^\dagger).
\end{equation}
They satisfy $U_f^3=U_{d_1}^2=1$ and $U_{d_1} U_f U_{d_1}=U_f^{\dagger}$. The three $\mathbb Z_2$ reflection symmetries are
\[
U_{d_1},\qquad U_{d_2}=U_f U_{d_1},\qquad U_{d_3}=U_{f}^\dagger U_{d_1}.
\]
The pair $(\sigma_j,\sigma_j^\dagger)$ transforms as the two-dimensional symmetry charge of $S_3$. Its two components carry conjugate $\mathbb Z_3$ charges and are exchanged by $U_{d_1}$. The symmetry fluxes correspond to ground states of Hamiltonians twisted by elements of $S_3$. The reflection flux has three components corresponding to twists by $U_{d_1},U_{d_2},U_{d_3}$, while the rotation flux has two components corresponding to twists by $U_f$, $U_{f}^\dagger$.

For $J>f$, the $\mathbb Z_3$ symmetry is spontaneously broken, and the three ordered ground states are permuted by $U_f$. Choosing the ground state with
\[
\langle\sigma_j\rangle=\langle\sigma_j^\dagger\rangle=v>0
\]
leaves $U_{d_1}$ unbroken. The two-dimensional $S_3$ symmetry charge therefore condenses in the equal combination of its two components, while the reflection flux multiplet condenses in the $U_{d_1}$ component.
For $J<f$, the full $S_3$ symmetry is unbroken, so no symmetry charge is condensed. The reflection and rotation fluxes condense in the equal-weight superposition of their components.

At the $J=f$ Potts critical point, the two phases share no nontrivial condensed defect: the reflection flux condenses on both sides, but in different components. Thus $K$ contains only the trivial defect, and the criterion predicts that every nontrivial topological defect becomes gapless and this is exactly what happens at the 3-state Potts critical point. The order parameter $\sigma_j$ flows to the Potts spin field with scaling dimension $\Delta=2/15$. The charge-conjugation-odd local operator flows to the chiral field with $\Delta=9/5$ and conformal spin $\pm1$ \cite{Mong2014parafermion}. The rotation flux defect flows to the nonlocal disorder field in the Potts CFT that is dual to the spin field. In the $D(S_3)$ SymTFT whose anyons are labeled as $A,B,\dots,H$, the two gapped phases have physical boundaries $A+C+D$ and $A+D+F$. The three-dimensional anyon $D$ corresponding to the reflection flux is condensed in different dimensions in these Lagrangian algebras, and therefore the condensable algebra at the transition is the trivial one, $\cK=A$. 

Another example is the transition from the fully $S_3$-broken phase to the phase in which $S_3$ is broken to $\mathbb{Z}_3$. The common condensate $K$ contains the one-dimensional symmetry charge of $S_3$ that is odd under reflections. This yields two critical ground states exchanged by charge conjugation, which realize chiral $\mathbb Z_3$ transitions with opposite chiralities \cite{Ravindran2026landau}. Both components of the two-dimensional $S_3$ symmetry charge are mutually local with $K$ and flow to the conjugate spin fields of the chiral clock theory. The rotation fluxes likewise become gapless and flow to nonlocal disorder fields, whereas the reflection fluxes act as $-1$ on the condensed charge and therefore remain gapped.
In the SymTFT, this transition is described by the condensable algebra $\cK=A+B$ between the gapped phases with physical boundaries $A+B+2C$ and $A+B+2F$. 

\subsection{Criterion for critical point and loop excitations in \texorpdfstring{$2+1$D}{2+1D}}\label{sec:criterion2d}

In $2+1$ dimensions, topological defects can be $0+1$D point defects or $1+1$D line defects. A nontrivial line defect is one that cannot be created by a $1$D finite-depth circuit but has to be created by a $1$D sequential circuit or a $2$D circuit. The ones created by a 1D sequential circuit are descendant defects, while the ones created by a 2D circuit are non-descendant. Descendant defects are gapped one-dimensional states formed by point excitations. A Cheshire string, along which an anyon condenses, is a basic example \cite{Alford1990,Preskill1990,Tantivasadakarn2024}. In our following discussion, it is important to take into account the 2-category structure of the defects. The line excitations can be interpreted as objects in a fusion 2-category, while the point excitations are morphisms, or domain walls, between them. Only when a line defect is condensed can we interpret its end point as an isolated point defect. Otherwise, it is not well defined to treat their end points as point defects. 

We will focus on $2+1$D anyon condensation transitions out of a topological phase and discuss the behavior of both line defects and point defects (when they are well defined) across the transition.  In a $2+1$D topologically ordered phase, closed anyon string operators realize trivial line defects. These anyon loops are condensed because the energy splitting between states differing by a noncontractible loop of the anyon string around the torus is exponentially small in the system size, while contractible loops act trivially on the ground state. Note that while the anyons have nontrivial braiding with each other, the closed (contractible) anyon loops all commute with each other and condense together in the topological phase. Moreover, since the loops are condensed, we can study their end points as point defects. Open anyon strings create nontrivial point excitations on top of these trivial line excitations. A topological phase can also have nontrivial line defects. For example, the toric code has five nontrivial line defects \cite{Kong2022topological, Zhao2025fusion}, which include the Cheshire strings.

An anyon condensation transition occurs when a collection of pointlike anyons becomes indistinguishable from the vacuum. The condensing anyons must be bosons, and the fusion channels selected by the condensate must close and remain unchanged when the anyons are exchanged. These data define a condensable algebra $\cA$ in the modular tensor category $\cC$ describing the original topological order. Across the transition, anyons may split and the split components may become condensed, confined, or remain deconfined \cite{Bais2009condensate}. The deconfined excitations of the new topological phase form the category of local $\cA$-modules, $\cC_{\cA}^{\mathrm{loc}}$ \cite{Kong2014anyon}. 

Microscopically, anyon condensation transitions can be implemented by adding additional degrees of freedom and local terms to string-net models that lower the energy of the condensing anyon \cite{Christian2023lattice, Lin2023}. Another route is via deformed wavefunctions \cite{Castelnovo2008topological}, whose correlations are governed by $2$D CFTs and not $3$D CFTs, and whose parent Hamiltonians have $z > 1$ quantum critical points. We focus on the first route via locally deformed Hamiltonians and propose the following criterion for the gapless excitations at the transition point. Since all anyons are gapped before the transition, only loop defects can stay condensed throughout the transition. The analogue to the collection of condensed topological defects $K$ in the $1+1$D criteria \ref{criterion1d} will be the set of condensed anyon loops. When an anyon component is either condensed or deconfined after the transition, their corresponding loop operator remains condensed. When an anyon component is confined, their loop operator becomes gapped.

\begin{criterion}\label{criterion2d}
    Consider a direct continuous anyon condensation transition out of a topological order $\cC$, driven by the condensable algebra $\cA$. Then:
    \begin{enumerate}
        \item Loops of anyons that split into only deconfined or condensed components remain condensed across the transition and form the set $K$. On a spatial torus, this means that the energy splitting between sectors with and without a noncontractible loop vanishes exponentially with the linear system size $L$. They correspond to the $a$ defects in Fig.~\ref{fig:defects}.
        
        \item All other anyon loops (of anyons that are at least partially confined) commute with the condensed loop set and become critical. On a spatial torus, this means that inserting a noncontractible loop leads to an energy splitting that vanishes algebraically with the linear system size $L$, as $1/L$ for a 3D CFT. After the transition, the loops become gapped. Therefore, they correspond to the $c$ defects in Fig.~\ref{fig:defects}.
        
        For anyon loops in the condensed set $K$, we can talk about the behavior of their end points -- the anyons across the transition.
      
        \item 
        A point-like anyon becomes critical if it braids trivially with every condensed loop in set $K$; its excitation gap then closes algebraically. 
        \item All other anyonic point excitations remain gapped. This happens when the anyon remains deconfined.
    \end{enumerate}
\end{criterion}

The point-like anyons that become fully condensed  (corresponding to the $d$ defect in Fig.~\ref{fig:defects}) or partially condensed and partially deconfined after the transition are critical at the transition. The ones that become fully deconfined (like the $e$ and $f$ defects in Fig.~\ref{fig:defects}) can be critical or gapped at the transition. When we discuss the commutation between anyon loops, we consider only contractible ones.

For a more mathematical perspective, let $\cB=\langle\cA\rangle\subset\cC$ be the fusion subcategory generated by the simple objects of the condensable algebra $\cA$. The loops condensed on both sides of the transition live in the M\"uger centralizer $\cB'_{\cC}$ (also denoted as $\cB'$), consisting of the anyons with trivial mutual braiding with every object of $\cB$. These loops in the M\"uger centralizer $\cB'_{\cC}$ act trivially at the transition -- inserting these loops along a noncontractible cycle induces a change in ground state energy that is at most exponentially small. In \cite{Cheng2026proliferation}, these loops are thought of as non-invertible 1-form symmetry operators and the symmetry they generate survives at the transition.

Critical point excitations can exist at the end points of condensed loops and must braid trivially with every condensed loop. They are therefore represented by objects in the M\"uger center $Z_2(\cB'_{\cC})$.  Since $\cC$ is modular, the double-centralizer theorem gives $(\cB'_{\cC})'_{\cC}=\cB$ \cite{Muger2003modular}, and hence
\begin{equation}
Z_2(\cB'_{\cC})
=\cB'_{\cC}\cap(\cB'_{\cC})'_{\cC}
=\cB'_{\cC}\cap\cB
=Z_2(\cB).
\end{equation}
In \cite{Cheng2026proliferation} and \cite{Schafer-Nameki2026proliferation}, the M\"uger center $Z_2(\cB)$ is used to construct a $3+1$D SymTFT bulk such that the critical point excitations at the $2+1$D transition become deconfined topological excitations in the $3+1$D bulk. But when $Z_2(\cB)$ is trivial, the bulk becomes trivial and does not capture the critical loop excitations at the $2+1$D transition.

\subsection{Examples in \texorpdfstring{$2+1$D}{2+1D}}\label{sec:criterionexamples}

\subsubsection{Toric code}

We start with the simplest example: the transition from toric code topological order $\cC=\{1,e,m,f\}$ to the trivial phase, driven by the condensation of $\cA=1+e$. In this case, $\cB=\langle \cA \rangle = \{1,e\}$ and $\cB'_\cC=\{1,e\}$. Our criterion therefore predicts that the $e$ point excitations and the $m$ and $f$ loop excitations become critical. Since the $e$ loop remains condensed, the $m$ and $f$ loops differ only by fusion with a condensed loop and flow to the same defect in the infrared. 

The lattice realization of this transition is the toric code in a magnetic field,
\begin{equation*}
    H = - J_v \sum_v A_v - J_p \sum_p B_p - h_x \sum_e \sigma^x_e - h_z \sum_e \sigma^z_e, 
\end{equation*}
where the plaquette and vertex terms are defined as 
\[
    A_v = \prod_{e \in v} \sigma^z_e, \qquad B_p = \prod_{e \in p} \sigma^x_e.
\]
The vertex and plaquette couplings are fixed to $J_v=J_p=1$ for all calculations.
An open $\sigma^x$ string creates a pair of $e$ particles, detected by $A_v=-1$ at its endpoints. We first consider the case $h_z=0$, so the field $h_x$ drives the condensation of $e$ anyons. Since every $B_p$ commutes with the Hamiltonian at $h_z=0$, i.e. the $e$ loop is condensed, the low-energy theory can be restricted to the flux-free sector $B_p=1$, where the model maps exactly to the transverse field Ising model \cite{hamma2008adiabatic, Werner2007breakdown},
\[
H_{\mathrm{Ising}} = -J_v\sum_{v}\tau^x_{v}
-h_x\sum_{\langle v,w\rangle} \tau^z_{v}\tau^z_{w},
\]
restricted to the even sector $\prod_{v}\tau^x_{v}=1$. For the square lattice, the transition occurs at $h_{x,c}/J_v \approx 0.328$ \cite{Blote2002cluster} and is described by the 3D Ising CFT (more precisely, by the gauged theory called Ising$^*$\cite{Schuler2016torusspectra}).

Under this mapping, the endpoint of an open $e$ string is represented by the Ising spin $\tau^z$ and therefore flows to the spin field $\sigma$ in the CFT. 

To diagnose the criticality of a noncontractible $m$ loop $W_m(C)$, consider the finite-size spectrum on a spatial torus. The eigenvalues of the two noncontractible $e$ loops label the four topological sectors of the toric code. These map to periodic or antiperiodic boundary conditions of the transverse field Ising model. Since $W_m(C)$ anticommutes with the $e$ loop intersecting $C$, inserting an $m$ loop changes one boundary condition, say from $(P,P)$ to $(A,P)$. Its energy cost is therefore the difference between the twisted and untwisted sector ground state energies of the Ising model,
\begin{equation}
\Delta_m(L)
=E_0^{(A,P)}(L)-E_0^{(P,P)}(L)
=\frac{v}{L}\,\varepsilon_m+o(L^{-1}),
\end{equation}
where $v$ is nonuniversal, while $\varepsilon_m$ is a universal number fixed by the CFT and the shape of the torus \cite{Schuler2016torusspectra,verresen2021gapless}. 

At fixed, sufficiently small $h_z>0$, increasing $h_x$ still drives an Ising$^*$ transition from the toric code to the trivial phase, with a critical field close to $h_x\approx0.33$. Our criterion predicts that the $m$ loop becomes critical, even though the bare $e$ loops no longer commute with the Hamiltonian. Consequently, their eigenvalues no longer define exact symmetry sectors whose ground state energies we can compare.

Nevertheless, the four lowest-lying states in the toric code phase with $h_z<h_x$ can still be characterized by the expectation values of the $e$ loops and their product, whose magnitudes are reduced by perimeter-law decay. In particular, the lowest energy state and third excited state (within the ground space) have $\langle W_e^x\rangle=\langle W_e^y\rangle\approx+1$ and $-1$, respectively, on small tori. The first two excited states within the ground space are approximately symmetric and antisymmetric combinations of states with $e$-loop expectation values close to $(+1,-1)$ and $(-1,+1)$, giving nearly zero individual loop expectations and negative $\langle W_e^xW_e^y\rangle$. The antisymmetric state is odd under $90^\circ$ rotations. This supports interpreting the three excited states within the ground space as originating from states with one or two noncontractible $m$ loops inserted.

\begin{figure*}[ht]
    \centering
    \includegraphics[]{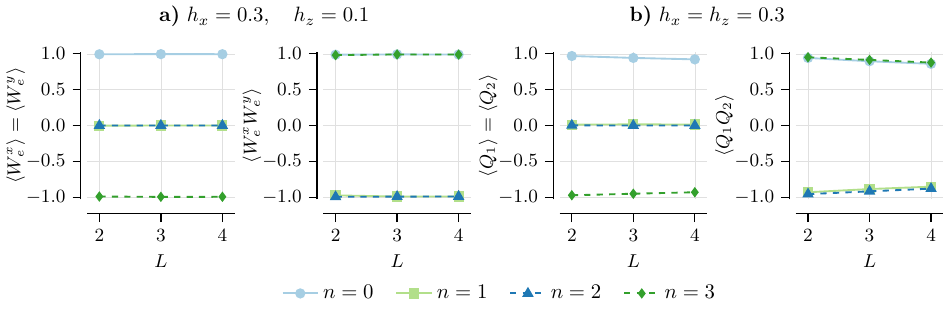}
    \caption{Expectation values of loop operators in the four lowest energy states $n=0,1,2,3$ in the toric code topological phase at a) $h_x=0.3, h_z=0.1$ and b) $h_x=h_z=0.3$. Markers show exact diagonalization data for $L=2,3,4$, lines are for guidance.}
    \label{fig:selfdualTCQ}
\end{figure*}

\begin{figure*}[ht]
    \centering
    \includegraphics[]{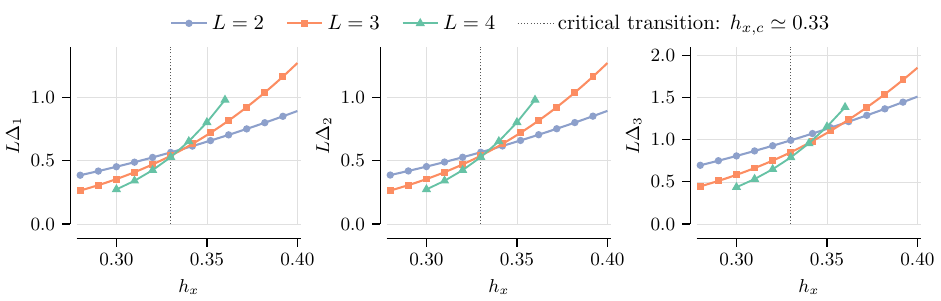}
    \caption{Scaled energy gaps $L\Delta_i$ of the three excited levels originating from the toric code ground states on $L\times L$ tori, at fixed $h_z=0.1$. Markers show exact diagonalization data for $L=2,3,4$, lines are for guidance.}
    \label{fig:TCgaps}
\end{figure*}

We illustrate the loop expectation values in Fig.~\ref{fig:selfdualTCQ}a) and plot the energy gaps across the transition in Fig.~\ref{fig:TCgaps}. Their size-dependent crossings move toward the critical transition at $h_{x,c} \approx 0.33$ as $L$ increases, consistent with asymptotic $1/L$ scaling.
The ED calculations were performed with QuSpin \cite{Quspin2017, Quspin2019}, and exploit the point group symmetries of the square lattice in the zero momentum sector.
The state associated with $\Delta_2$ is odd under $90^\circ$ rotations and diagonal reflections, and even under $180^\circ$ rotations and reflections about the coordinate axes. The states associated with $\Delta_1$ and $\Delta_3$ are fully symmetric.

\subsubsection{$D(\mathbb{Z}_N)$}

The anyons of the quantum double $D(\mathbb{Z}_N)$ are labeled by $e^a m^b$, with $a,b\in\mathbb{Z}_N$. Condensing the electric Lagrangian algebra $\cA=\oplus_{a=0}^{N-1} e^a$ drives the system into the trivial phase. According to our criterion, all $e$ loops are condensed, while all $m$ loops and $e$ point particles are critical. 

The transition can be mapped to a $\mathbb{Z}_N$ symmetry-breaking transition and is therefore described by the $N$-state clock model. For $N \geq 5$, the transition is continuous and in the $O(2)$ Wilson-Fisher universality class.
The torus spectrum of this CFT was studied in Ref.~\cite{Whitsitt2017wilsonfisher}. A non-contractible $m$ loop corresponds to a $\mathbb{Z}_N$ twisted boundary condition, whose ground state energy differs from that of the untwisted model by $\mathcal{O}(1/L)$.

\subsubsection{Doubled Ising}

We now apply our criterion to the transition from the doubled Ising to the trivial phase discussed in Section~\ref{sec:DIs}. The condensable algebra $\cA_2=1 \bar 1 + \psi \bar \psi + \sigma \bar \sigma$ generates the subcategory $\cB=\{1 \bar 1, \psi \bar \psi, \psi \bar 1, 1 \bar \psi, \sigma \bar \sigma \}$. The anyon loops in $\cB'=\{1 \bar 1, \psi \bar \psi\}$ should remain condensed, while all other loops become critical. The point excitation corresponding to $\psi \bar \psi$ is also predicted to become critical.

The doubled Ising phase can be obtained by gauging the global $\mathbb{Z}_2$ symmetry exchanging $e$ and $m$ in the toric code \cite{Barkeshli2014fractionalization}. Applying the gauging across the self-dual toric code transition relates it to the doubled Ising to trivial phase transition \cite{Ji2026selfdual}. 
Numerical studies support a continuous self-dual toric code transition at around $h_x=h_z\approx 0.34$ \cite{Wu2012toriccode, Nahum2021selfdual}. 

The $\sigma \bar \sigma$ anyon splits into $e \oplus m$ under de-equivariantization (ungauging), so its predicted critical loop behavior can be investigated through the $e$ and $m$ loops of the self-dual toric code. Similarly, $\psi\bar 1$ and $1\bar \psi$ both map to the toric code anyon $f$. The $\sigma \bar 1$ and $1 \bar \sigma$ anyons instead map to twist defects of the $e$-$m$ exchange symmetry \cite{Bombin2010twist}. Their loops correspond to dislocation defect lines wrapping around a noncontractible cycle of the torus. The $\psi \bar \psi$ anyon maps to the trivial toric code anyon, but we can probe the corresponding point excitation through an excitation that is odd under the $e$-$m$ exchange symmetry. The expectation value of the corresponding edge order parameter $\sigma^x_e-\sigma^z_e$ across the transition was computed with iPEPS in Ref.~\cite{Xu2025fredenhagen}. After extrapolating in inverse bond dimension, they found its critical scaling consistent with the field $\phi_x^2-\phi_z^2$ in the 3D $XY^*$ field theory.

Because the only nontrivial condensed loop, $\psi\bar\psi$, braids trivially with $\sigma\bar\sigma$, it cannot distinguish states with and without an inserted $\sigma\bar\sigma$ loop. This differs from the toric code at $h_z=0$, where the eigenvalues of the condensed $e$ loops detect inserted $m$ loops. 

We follow an approach similar to that used for the toric code transition at fixed $h_z$ ($0<h_z<h_x$). Instead of the individual $e$ loops, we consider 
\[
    Q_1 = \frac{1}{\sqrt{2}} (W^x_e + W^y_m), \qquad Q_2 = \frac{1}{\sqrt{2}} (W^y_e + W^x_m).
\]
Note that $Q_1^2=Q_2^2=\mathbb{I}$ and $[Q_1,Q_2]=0$. 
In the topological phase along $h_x=h_z$, among the four states in the ground space,
the lowest energy state and the third excited state have $\langle Q_1\rangle=\langle Q_2\rangle\approx+1$ and $-1$, respectively, on small tori, with magnitudes reduced by perimeter law decay. The first two excited states are approximately symmetric and antisymmetric combinations of states with opposite $Q_1$ and $Q_2$ values, giving nearly zero individual expectations of $Q_1$ and $Q_2$ and negative $\langle Q_1Q_2\rangle$. These observations support interpreting the three excited states as due to combinations of $e$- and $m$-loop insertions. 
More explicitly, $W_e^x W_m^y$ anticommutes with $Q_1$ and commutes with $Q_2$, while $W_e^y W_m^x$ does the reverse. Therefore, the first and second excited states correspond to the symmetric and antisymmetric combinations of these two insertions acting on the ground state. The operator $W_f^xW_f^y$ anticommutes with both $Q_1$ and $Q_2$, so the third excited state corresponds to $f$-loop insertions along both directions.

\begin{figure*}[ht]
    \centering
    \includegraphics[]{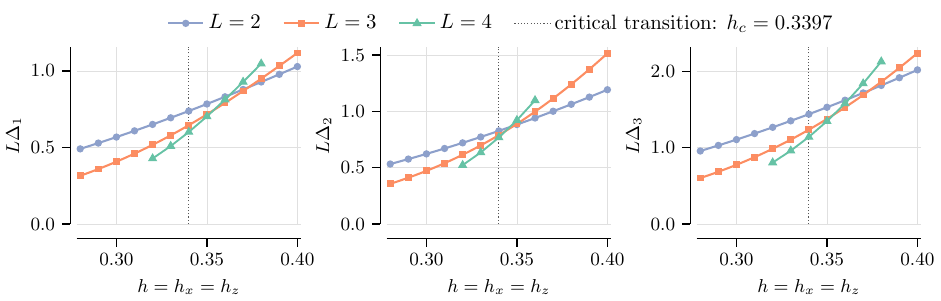}
    \caption{Scaled energy gaps $L\Delta_i$ of the three excited levels originating from the toric code ground states on $L\times L$ tori, along the self-dual line $h_x=h_z=h$. Markers show exact diagonalization data for $L=2,3,4$, lines are for guidance.}
    \label{fig:selfdualTCtransition}
\end{figure*}

We plot the loop expectation values in Fig.~\ref{fig:selfdualTCQ}b) and the rescaled excitation gaps $L\Delta_i=L(E_i-E_0)$, $i=1,2,3$, in Fig.~\ref{fig:selfdualTCtransition}. Their size-dependent crossings move toward the self-dual transition as $L$ increases, consistent with $1/L$ scaling at criticality. However, the accessible sizes $L=2,3,4$ exhibit substantial finite-size effects. Larger system calculations, potentially using quantum Monte Carlo, are needed to establish the asymptotic scaling.

\subsubsection{Doubled Fibonacci}

Next we consider the transition from the doubled Fibonacci phase $\cC = Z(\mathrm{Fib}) = \{1,\tau,\bar \tau, \tau \bar \tau\}$ to the trivial phase, driven by the condensation of $\cA=1+\tau \bar \tau$. Since the fusion subcategory $\cB$ generated by the algebra $\cA$ is the whole category, $\cB=\cC$ and $\cB'=\{1\}$, there are no condensed loops and our criterion predicts that all nontrivial loops $\{\tau, \bar \tau, \tau \bar \tau\}$ become critical. 

A lattice realization of this transition is the Levin-Wen string-net \cite{Levin2004stringnet} constructed from the Fibonacci category, together with a string tension term,
\[
    H = - \sum_p B_p - \lambda \sum_\ell L_\ell.
\]
Here $B_p$ projects onto trivial flux through the plaquette $p$, while $L_\ell$ projects the edge $\ell$ onto the vacuum label. We work in the subspace satisfying the Fibonacci branching rules, so the vertex projectors are fixed to $A_v=1$ and have been omitted.

An early study based on series expansions and exact diagonalization found a continuous transition between the doubled Fibonacci and trivial phases \cite{Schulz2013fibonacci}. 
A subsequent variational mean-field analysis \cite{Dusuel2015meanfield} instead predicted a first order transition, in agreement with later tensor network calculations \cite{Schotte2019tensor}. This perturbed string-net model therefore does not provide a suitable test of our criterion for continuous transitions.

\subsubsection{$D(S_3)$}

Finally we apply our criteria to anyon condensation transitions in the quantum double $\cC=D(S_3)$, which we discussed in the field theory framework in Section~\ref{sec:DS3}. 
Consider the algebra $\cA_2=A+C$ which drives a transition into the toric code phase. Since the anyon $B$ appears in the fusion product of $C$, $\cB=\{A,B,C\}$, and $\cB'=\cB$. Therefore, the $B$ and $C$ loops are condensed, and the corresponding point excitations become critical. The remaining loops labeled by $D,E,\dots,H$ become critical as well. 
Recent field-theory analyses argue that the generic 2+1D transition is a gauged 3D three-state Potts transition, which is weakly first order \cite{Schafer-Nameki2026proliferation,Cheng2026proliferation,Chavda2026orderparam}.

Condensing the Lagrangian algebra $\cA_3=A+B+2C$ instead drives the system into the trivial phase. Here $\cB=\cB'=\{A,B,C\}$ as well, so the set of critical loops and anyons is the same as above. Since $\cA$ is equivalent to Rep$(G)$ here with $G=S_3$, the transition can be mapped to an $S_3$ symmetry breaking transition, which is generally first order \cite{Schafer-Nameki2026proliferation, Cheng2026proliferation}, as we also find in our field theory procedure. 
A microscopic realization was recently studied using Monte Carlo simulations \cite{Flores-Calderon2026condensation}. The system underwent a first order transition into the trivial phase, at which the order parameters for both the $B$ and $C$ open strings jumped discontinuously.

The trivial phase can also be reached by condensing the Lagrangian algebra $\cA_6=A+C+D$. 
The subcategory generated by $\cA$ is now the whole anyon theory, $\cB=D(S_3)$. Therefore, no loops are condensed, and all become critical. 
Because this algebra is not equivalent to Rep$(G)$, the transition cannot simply be mapped to a $G$ symmetry breaking transition. It can be regarded as a two-step process, first condensing $A+C$ as above, and then condensing $A+D'$, where $D'$ is the deconfined part of $D$ after the first step; see also our discussion in Section~\ref{sec:DS3}. If we assume that the direct transition driven by $A+C+D$ can be obtained by combining the transition driven by $A+C$ and that driven by $A+D'$, we see indeed all the loops are critical. In the first step of the transition, $B$ and $C$ are critical point excitations while $D$--$H$ are critical loop excitations. At the first transition, $C$ splits into two parts, one gets condensed while the other remains deconfined and is identified with $B$. $D$ also splits into two parts, one part remains deconfined while the other becomes confined. In the second step of the transition, $D'$, which is now a $\mathbb{Z}_2$ abelian anyon, is condensed. At this critical point, $D'$ is the critical point excitation and $B$ is the critical loop excitation. When combined with the first step of the transition, we see that all loop excitations are critical.

The closely related $A+D+F$ condensation transition was studied with Monte Carlo simulations in Ref.~\cite{Flores-Calderon2026condensation}, and found to be first order as well.

\section{Discussion}\label{sec:discussion}

In this work, we explored how the categorical data of anyon condensation constrains the corresponding phase transition. When the condensing anyons have integer quantum dimensions, we can assign fields to their internal degrees of freedom and construct interaction terms of a Ginzburg-Landau field theory from fusion trees into the vacuum.
As a separate approach, we identified the loop excitations of the topological order that stay condensed at the transition and argued that all other loops should become critical. Point excitations associated with condensed loops are well defined. Among them, those that braid trivially with every condensed loop are expected to become critical. We discussed how this criterion is realized for the toric code in a magnetic field, and stated our expectations for other transitions. 

Even though the anyon condensation transition can be thought of as purely driven by point-like excitations, the critical excitations at the transition necessarily involve line-like objects as well due to their nontrivial braiding relation with the point-like excitations. A perspective we try to promote in this paper is that, when thinking about topological defects in a $2+1$D critical theory, we need to start with their 2-category structure. That is, the objects at the first level are line-like defects and the point-like defects come in only as the domain wall (morphism) between line-like defects. This perspective is inspired by the 2-category structure of topological excitations in gapped $3+1$D topological phases\cite{Johnson-Freyd2022,Kong2020classification,Kong2014arxiv} and the success of the holographic principle in studying $1+1$D phase transitions using the $2+1$D SymTFT formalism.

A complete 2-category structure of topological defects should involve not just loop-like defects and point-like defects, but also line-like defects that are composed of point-like defects along the line. Such defects are called `descendant' defects and examples include the `Cheshire' string\cite{Alford1990,Preskill1990,Tantivasadakarn2024}. We did not discuss such defects in this paper, therefore our analysis is a very incomplete attempt to use the 2-category structure to study topological defects.

An inconvenient fact is that the 2-category structure of topological defects in $2+1$D is not very compatible with the requirement of conformal symmetry. Introduction of line-like defects breaks the conformal symmetry to a lower dimensional one. Moreover, if one tries to probe the effect of nontrivial topology of the line-like defects by putting the system on a spatial torus, the system is no longer conformally equivalent to infinite space and the operator-state correspondence is lost. If we insist on having the conformal symmetry, we need to put the system on a sphere and lose access to non-contractible loops. 

Anyon condensation transitions are a special class of $2+1$D transitions. While the anyons exist at the end of string operators, the strings are invisible before the transition. Therefore, the transition is driven by the fluctuation of $0+1$D point-like excitations -- the condensing anyons. Because of this, we can modify the conventional Ginzburg-Landau theory in non-dramatic ways to formulate a description of anyon condensation transitions. Generic $2+1$D transitions, however, are driven by the fluctuation of $1+1$D line-like excitations. For example, the transition between symmetry-protected topological phases and that between the chiral and non-chiral superconductors\cite{Ji2024topological}. A proper description of such transitions would need to capture the fluctuation of line-like excitations and results in more `stringy' type field theories\cite{Iqbal2022}.


\begin{acknowledgments}
We are grateful for inspiring discussions with Vibhu Ravindran and Dachuan Lu. X.C. is supported by the Simons collaboration on ``Ultra-Quantum Matter'' (grant number 651438), the Simons Investigator Award (award ID 828078), the Institute for Quantum Information and Matter at Caltech (grant number PHY-2317110), the Walter Burke Institute for Theoretical Physics at Caltech and the Leinweber Forum for Theoretical Physics at Caltech. L.E. is supported by the Walter Burke Institute for Theoretical Physics at Caltech and the Simons collaboration on `Ultra-Quantum Matter'' (grant number 651438). When finishing this work, we became aware of related work by Sakura Sch\"afer-Nameki and Yuhan Gai on anyon proliferation transitions in Quantum Double models.
\end{acknowledgments}

\bibliography{references}

@misc{Ravindran2026landau,
	archiveprefix = {arXiv},
	author = {Ravindran, Vibhu and Eck, Luisa and Chen, Xie},
	eprint = {2609.06751},
	primaryclass = {cond-mat.str-el},
	title = {{Ginzburg-Landau Theory for Non-Invertible Symmetry-Breaking Transitions}},
	year = {2026}}

@misc{Cheng2026proliferation,
	archiveprefix = {arXiv},
	author = {Cheng, Meng},
	eprint = {2608.20604},
	primaryclass = {cond-mat.str-el},
	title = {{Note on proliferation transitions of non-Abelian anyons}},
	year = {2026}}

@misc{Schafer-Nameki2026proliferation,
	archiveprefix = {arXiv},
	author = {Sch{\"a}fer-Nameki, Sakura and Zheng, Yunqin and Antinucci, Andrea},
	eprint = {2608.12303},
	primaryclass = {cond-mat.str-el},
	title = {{Proliferation Transitions for Non-Abelian Anyons}},
	year = {2026}}

@article{Cheng2026abelianproliferation,
	author = {Cheng, Meng and Seiberg, Nathan},
	doi = {10.1103/bfzc-7lm2},
	issue = {5},
	journal = {Phys. Rev. B},
	month = {Jul},
	numpages = {26},
	pages = {055103},
	publisher = {American Physical Society},
	title = {Proliferation transitions from a topological phase in $2+1$ dimensions},
	url = {https://link.aps.org/doi/10.1103/bfzc-7lm2},
	volume = {114},
	year = {2026}}

@misc{Ji2024topological,
	archiveprefix = {arXiv},
	author = {Ji, Wenjie and Chen, Xie},
	eprint = {2407.02488},
	primaryclass = {cond-mat.str-el},
	title = {{Topological defects of 2+1D systems from line excitations in 3+1D bulk}},
	year = {2024}}

@misc{Ji2025bulk,
	author = {Ji, Wenjie and Stephen, David T and Levin, Michael and Chen, Xie},
	eprint        = {2506.11288},
    archivePrefix = {arXiv},
	title = {Bulk excitations of invertible phases},
	year = {2025}}

@article{Werner2007breakdown,
	author = {Trebst, Simon and Werner, Philipp and Troyer, Matthias and Shtengel, Kirill and Nayak, Chetan},
	doi = {10.1103/PhysRevLett.98.070602},
	issue = {7},
	journal = {Phys. Rev. Lett.},
	month = {Feb},
	numpages = {4},
	pages = {070602},
	publisher = {American Physical Society},
	title = {Breakdown of a Topological Phase: Quantum Phase Transition in a Loop Gas Model with Tension},
	url = {https://link.aps.org/doi/10.1103/PhysRevLett.98.070602},
	volume = {98},
	year = {2007}}

@article{hamma2008adiabatic,
	author = {Hamma, Alioscia and Lidar, Daniel A.},
	doi = {10.1103/PhysRevLett.100.030502},
	issue = {3},
	journal = {Phys. Rev. Lett.},
	month = {Jan},
	numpages = {4},
	pages = {030502},
	publisher = {American Physical Society},
	title = {Adiabatic Preparation of Topological Order},
	url = {https://link.aps.org/doi/10.1103/PhysRevLett.100.030502},
	volume = {100},
	year = {2008}}

@article{Blote2002cluster,
	author = {Bl\"ote, Henk W. J. and Deng, Youjin},
	doi = {10.1103/PhysRevE.66.066110},
	issue = {6},
	journal = {Phys. Rev. E},
	month = {Dec},
	numpages = {8},
	pages = {066110},
	publisher = {American Physical Society},
	title = {Cluster Monte Carlo simulation of the transverse Ising model},
	url = {https://link.aps.org/doi/10.1103/PhysRevE.66.066110},
	volume = {66},
	year = {2002}}

@article{verresen2021gapless,
	author = {Verresen, Ruben and Thorngren, Ryan and Jones, Nick G. and Pollmann, Frank},
	doi = {10.1103/PhysRevX.11.041059},
	issue = {4},
	journal = {Phys. Rev. X},
	month = {Dec},
	numpages = {41},
	pages = {041059},
	publisher = {American Physical Society},
	title = {Gapless Topological Phases and Symmetry-Enriched Quantum Criticality},
	url = {https://link.aps.org/doi/10.1103/PhysRevX.11.041059},
	volume = {11},
	year = {2021}}

@article{Dusuel2015meanfield,
	author = {Dusuel, S\'ebastien and Vidal, Julien},
	doi = {10.1103/PhysRevB.92.125150},
	issue = {12},
	journal = {Phys. Rev. B},
	month = {Sep},
	numpages = {10},
	pages = {125150},
	publisher = {American Physical Society},
	title = {Mean-field ansatz for topological phases with string tension},
	url = {https://link.aps.org/doi/10.1103/PhysRevB.92.125150},
	volume = {92},
	year = {2015}}

@article{Levin2004stringnet,
	author = {Levin, Michael A. and Wen, Xiao-Gang},
	doi = {10.1103/PhysRevB.71.045110},
	issue = {4},
	journal = {Phys. Rev. B},
	month = {Jan},
	numpages = {21},
	pages = {045110},
	publisher = {American Physical Society},
	title = {String-net condensation: A physical mechanism for topological phases},
	url = {https://link.aps.org/doi/10.1103/PhysRevB.71.045110},
	volume = {71},
	year = {2005}}

@article{Schuler2016torusspectra,
	author = {Schuler, Michael and Whitsitt, Seth and Henry, Louis-Paul and Sachdev, Subir and L\"auchli, Andreas M.},
	doi = {10.1103/PhysRevLett.117.210401},
	issue = {21},
	journal = {Phys. Rev. Lett.},
	month = {Nov},
	numpages = {6},
	pages = {210401},
	publisher = {American Physical Society},
	title = {Universal Signatures of Quantum Critical Points from Finite-Size Torus Spectra: A Window into the Operator Content of Higher-Dimensional Conformal Field Theories},
	url = {https://link.aps.org/doi/10.1103/PhysRevLett.117.210401},
	volume = {117},
	year = {2016}}

@article{Bhardwaj2025hasse,
	author = {Bhardwaj, Lakshya and Pajer, Daniel and Sch{\"a}fer-Nameki, Sakura and Warman, Alison},
	doi = {10.21468/SciPostPhys.19.4.113},
	journal = {SciPost Phys.},
	pages = {113},
	publisher = {SciPost},
	title = {Hasse diagrams for gapless SPT and SSB phases with non-invertible symmetries},
	url = {https://scipost.org/10.21468/SciPostPhys.19.4.113},
	volume = {19},
	year = {2025}}

@article{Mong2014parafermion,
	archiveprefix = {arXiv},
	author = {Mong, Roger S. K. and Clarke, David J. and Alicea, Jason and Lindner, Netanel H. and Fendley, Paul},
	doi = {10.1088/1751-8113/47/45/452001},
	eprint = {1406.0846},
	journal = {J. Phys. A},
	number = {45},
	pages = {452001},
	primaryclass = {cond-mat.stat-mech},
	title = {{Parafermionic conformal field theory on the lattice}},
	volume = {47},
	year = {2014}}

@article{Bais2009condensate,
	author = {Bais, F. A. and Slingerland, J. K.},
	doi = {10.1103/PhysRevB.79.045316},
	issue = {4},
	journal = {Phys. Rev. B},
	month = {Jan},
	numpages = {27},
	pages = {045316},
	publisher = {American Physical Society},
	title = {Condensate-induced transitions between topologically ordered phases},
	url = {https://link.aps.org/doi/10.1103/PhysRevB.79.045316},
	volume = {79},
	year = {2009}}

@article{Kong2014anyon,
	author = {Kong, Liang},
	journal = {Nuclear Physics B},
	pages = {436--482},
	publisher = {Elsevier},
	title = {{Anyon condensation and tensor categories}},
	volume = {886},
	year = {2014}}

@article{Huse1982,
  title = {Domain Walls and the Melting of Commensurate Surface Phases},
  author = {Huse, David A. and Fisher, Michael E.},
  journal = {Phys. Rev. Lett.},
  volume = {49},
  issue = {11},
  pages = {793--796},
  numpages = {0},
  year = {1982},
  month = {Sep},
  publisher = {American Physical Society},
  doi = {10.1103/PhysRevLett.49.793},
  url = {https://link.aps.org/doi/10.1103/PhysRevLett.49.793}
}

@article{burnell2018anyon,
	author = {Burnell, Fiona J},
	journal = {Annual Review of Condensed Matter Physics},
	pages = {307--327},
	publisher = {Annual Reviews},
	title = {{Anyon condensation and its applications}},
	volume = {9},
	year = {2018}}

@article{Christian2023lattice,
	archiveprefix = {arXiv},
	author = {Christian, Jessica and Green, David and Huston, Peter and Penneys, David},
	doi = {10.1007/JHEP09(2023)055},
	eprint = {2303.04711},
	journal = {JHEP},
	pages = {055},
	primaryclass = {cond-mat.str-el},
	title = {{A lattice model for condensation in Levin-Wen systems}},
	volume = {09},
	year = {2023}}

@article{Lin2023,
	archiveprefix = {arXiv},
	author = {Lin, Chien-Hung and Burnell, Fiona J.},
	doi = {10.1103/PhysRevB.110.115127},
	eprint = {2303.07291},
	journal = {Phys. Rev. B},
	number = {11},
	pages = {115127},
	primaryclass = {cond-mat.str-el},
	title = {{Anyon condensation in string-net models}},
	volume = {110},
	year = {2024}}

@article{Zhao2024nonabelian,
	archiveprefix = {arXiv},
	author = {Zhao, Yu and Wan, Yidun},
	doi = {10.1007/JHEP05(2025)156},
	eprint = {2409.05852},
	journal = {JHEP},
	pages = {156},
	primaryclass = {cond-mat.str-el},
	title = {{Nonabelian anyon condensation in 2+1d topological orders: A string-net model realization}},
	volume = {05},
	year = {2025}}

@article{Schotte2019tensor,
	author = {Schotte, Alexis and Carrasco, Jose and Vanhecke, Bram and Vanderstraeten, Laurens and Haegeman, Jutho and Verstraete, Frank and Vidal, Julien},
	doi = {10.1103/PhysRevB.100.245125},
	issue = {24},
	journal = {Phys. Rev. B},
	month = {Dec},
	numpages = {8},
	pages = {245125},
	publisher = {American Physical Society},
	title = {Tensor-network approach to phase transitions in string-net models},
	url = {https://link.aps.org/doi/10.1103/PhysRevB.100.245125},
	volume = {100},
	year = {2019}}

@misc{Flores-Calderon2026condensation,
	archiveprefix = {arXiv},
	author = {Flores-Calder{\'o}n, Rafael and Pollmann, Frank and Knap, Michael},
	eprint = {2609.19282},
	primaryclass = {cond-mat.str-el},
	title = {{Non-Abelian Anyon Condensation: a Path-Integral Monte Carlo Approach}},
	year = {2026}}

@article{Ji2026selfdual,
	archiveprefix = {arXiv},
	author = {Ji, Wenjie and Lanzetta, Ryan A. and Zhou, Zheng and Wang, Chong},
	doi = {10.1103/flfk-m8lj},
	eprint = {2601.20945},
	journal = {Phys. Rev. Lett.},
	number = {8},
	pages = {086505},
	primaryclass = {cond-mat.str-el},
	title = {{Self-Dual Higgs Transitions: Toric Code and Beyond}},
	volume = {137},
	year = {2026}}

@article{Dumitrescu2026selfdual,
	author = {Dumitrescu, Thomas T and Niro, Pierluigi and Thorngren, Ryan},
	journal = {arXiv preprint arXiv:2602.23420},
	title = {From {$\mathrm{QED}_3$} to Self-Dual Multicriticality in the {Fradkin--Shenker} Model},
	year = {2026}}

@article{Schulz2013fibonacci,
	author = {Schulz, Marc Daniel and Dusuel, S\'ebastien and Schmidt, Kai Phillip and Vidal, Julien},
	doi = {10.1103/PhysRevLett.110.147203},
	issue = {14},
	journal = {Phys. Rev. Lett.},
	month = {Apr},
	numpages = {5},
	pages = {147203},
	publisher = {American Physical Society},
	title = {Topological Phase Transitions in the Golden String-Net Model},
	url = {https://link.aps.org/doi/10.1103/PhysRevLett.110.147203},
	volume = {110},
	year = {2013}}

@misc{Lu2026selfdual,
	archiveprefix = {arXiv},
	author = {Lu, Da-Chuan and Wang, Chong and Vishwanath, Ashvin},
	eprint = {2608.05294},
	primaryclass = {cond-mat.str-el},
	title = {Self-Dual {$S_3$} Gauge Theory in {$2+1$}D: Lattice Model and Topological Phase Transitions},
	year = {2026}}

@article{Bhardwaj2025,
	archiveprefix = {arXiv},
	author = {Bhardwaj, Lakshya and Bottini, Lea E. and Pajer, Daniel and Sch{\"a}fer-Nameki, Sakura},
	doi = {10.21468/SciPostPhys.18.1.032},
	eprint = {2310.03784},
	journal = {SciPost Phys.},
	number = {1},
	pages = {032},
	primaryclass = {hep-th},
	title = {{Gapped phases with non-invertible symmetries: (1+1)d}},
	volume = {18},
	year = {2025}}

@article{Ji2020,
	author = {Ji, Wenjie and Wen, Xiao-Gang},
	doi = {10.1103/PhysRevResearch.2.033417},
	issue = {3},
	journal = {Phys. Rev. Res.},
	month = {Sep},
	numpages = {24},
	pages = {033417},
	publisher = {American Physical Society},
	title = {Categorical symmetry and noninvertible anomaly in symmetry-breaking and topological phase transitions},
	url = {https://link.aps.org/doi/10.1103/PhysRevResearch.2.033417},
	volume = {2},
	year = {2020}}

@article{Castelnovo2008topological,
	author = {Castelnovo, Claudio and Chamon, Claudio},
	doi = {10.1103/PhysRevB.77.054433},
	issue = {5},
	journal = {Phys. Rev. B},
	month = {Feb},
	numpages = {14},
	pages = {054433},
	publisher = {American Physical Society},
	title = {Quantum topological phase transition at the microscopic level},
	url = {https://link.aps.org/doi/10.1103/PhysRevB.77.054433},
	volume = {77},
	year = {2008}}

@misc{Chavda2026orderparam,
	archiveprefix = {arXiv},
	author = {Chavda, Ameya and Delcamp, Clement and Turzillo, Alex and You, Minyoung},
	eprint = {2605.27193},
	primaryclass = {cond-mat.str-el},
	title = {{Algebras of order parameters in one-dimensional spin systems}},
	year = {2026}}

@article{Burnell2012,
	author = {Burnell, F J and Simon, Steven H and Slingerland, J K},
	doi = {10.1088/1367-2630/14/1/015004},
	journal = {New Journal of Physics},
	month = {jan},
	number = {1},
	pages = {015004},
	publisher = {IOP Publishing},
	title = {Phase transitions in topological lattice models via topological symmetry breaking},
	url = {https://doi.org/10.1088/1367-2630/14/1/015004},
	volume = {14},
	year = {2012}}

@article{Zhao2025fusion,
	author = {Zhao, Jingyu and Chen, Xie},
	doi = {10.21468/SciPostPhys.18.3.106},
	journal = {SciPost Phys.},
	pages = {106},
	publisher = {SciPost},
	title = {Fusion of low-entanglement excitations in 2D toric code},
	url = {https://scipost.org/10.21468/SciPostPhys.18.3.106},
	volume = {18},
	year = {2025}}

@misc{Kong2022topological,
	archiveprefix = {arXiv},
	author = {Kong, Liang and Zhang, Zhi-Hao},
	eprint = {2205.05565},
	primaryclass = {cond-mat.str-el},
	title = {{An invitation to topological orders and category theory}},
	year = {2022}}

@article{Muger2003modular,
	author = {M{\"u}ger, Michael},
	doi = {10.1112/S0024611503014187},
	journal = {Proceedings of the London Mathematical Society},
	number = {2},
	pages = {291--308},
	title = {On the Structure of Modular Categories},
	volume = {87},
	year = {2003}}

@article{Chatterjee2023holographic,
	author = {Chatterjee, Arkya and Wen, Xiao-Gang},
	doi = {10.1103/PhysRevB.108.075105},
	issue = {7},
	journal = {Phys. Rev. B},
	month = {Aug},
	numpages = {47},
	pages = {075105},
	publisher = {American Physical Society},
	title = {Holographic theory for continuous phase transitions: Emergence and symmetry protection of gaplessness},
	url = {https://link.aps.org/doi/10.1103/PhysRevB.108.075105},
	volume = {108},
	year = {2023}}

@misc{Garcia-Sepulveda:2026fud,
	archiveprefix = {arXiv},
	author = {Garc{\'\i}a-Sep{\'u}lveda, Diego and Lu, Da-Chuan},
	eprint = {2610.00452},
	primaryclass = {cond-mat.str-el},
	title = {{Anyon Proliferation and Anyon Superconductivity in Higgsing Transitions via Conformal Embeddings}},
	year = {2026}}

@article{Nahum2021selfdual,
	author = {Somoza, Andr{\'e}s M. and Serna, Pablo and Nahum, Adam},
	doi = {10.1103/PhysRevX.11.041008},
	issue = {4},
	journal = {Phys. Rev. X},
	month = {Oct},
	numpages = {32},
	pages = {041008},
	publisher = {American Physical Society},
	title = {Self-Dual Criticality in Three-Dimensional {$\mathbb{Z}_2$} Gauge Theory with Matter},
	url = {https://link.aps.org/doi/10.1103/PhysRevX.11.041008},
	volume = {11},
	year = {2021}}

@article{Wu2012toriccode,
	author = {Wu, Fengcheng and Deng, Youjin and Prokof'ev, Nikolay},
	doi = {10.1103/PhysRevB.85.195104},
	issue = {19},
	journal = {Phys. Rev. B},
	month = {May},
	numpages = {7},
	pages = {195104},
	publisher = {American Physical Society},
	title = {Phase diagram of the toric code model in a parallel magnetic field},
	url = {https://link.aps.org/doi/10.1103/PhysRevB.85.195104},
	volume = {85},
	year = {2012}}

@article{Bombin2010twist,
	author = {Bombin, H.},
	doi = {10.1103/PhysRevLett.105.030403},
	issue = {3},
	journal = {Phys. Rev. Lett.},
	month = {Jul},
	numpages = {4},
	pages = {030403},
	publisher = {American Physical Society},
	title = {Topological Order with a Twist: Ising Anyons from an Abelian Model},
	url = {https://link.aps.org/doi/10.1103/PhysRevLett.105.030403},
	volume = {105},
	year = {2010}}

@article{Barkeshli2014fractionalization,
	author = {Barkeshli, Maissam and Bonderson, Parsa and Cheng, Meng and Wang, Zhenghan},
	doi = {10.1103/PhysRevB.100.115147},
	issue = {11},
	journal = {Phys. Rev. B},
	month = {Sep},
	numpages = {99},
	pages = {115147},
	publisher = {American Physical Society},
	title = {Symmetry fractionalization, defects, and gauging of topological phases},
	url = {https://link.aps.org/doi/10.1103/PhysRevB.100.115147},
	volume = {100},
	year = {2019}}

@article{Xu2025fredenhagen,
	author = {Xu, Wen-Tao and Pollmann, Frank and Knap, Michael},
	journal = {npj Quantum Information},
	number = {1},
	pages = {74},
	publisher = {Nature Publishing Group UK London},
	title = {Critical behavior of Fredenhagen-Marcu string order parameters at topological phase transitions with emergent higher-form symmetries},
	volume = {11},
	year = {2025}}

@misc{Cheng2026peeling,
	archiveprefix = {arXiv},
	author = {Cheng, Meng and Seiberg, Nathan},
	eprint = {2610.04704},
	primaryclass = {cond-mat.str-el},
	title = {{Peeling off Abelian Anyons and Proliferation Transitions}},
	year = {2026}}

@article{Whitsitt2017wilsonfisher,
	author = {Whitsitt, Seth and Schuler, Michael and Henry, Louis-Paul and L\"auchli, Andreas M. and Sachdev, Subir},
	doi = {10.1103/PhysRevB.96.035142},
	issue = {3},
	journal = {Phys. Rev. B},
	month = {Jul},
	numpages = {23},
	pages = {035142},
	publisher = {American Physical Society},
	title = {Spectrum of the Wilson-Fisher conformal field theory on the torus},
	url = {https://link.aps.org/doi/10.1103/PhysRevB.96.035142},
	volume = {96},
	year = {2017}}

@misc{Kong2015,
	archiveprefix = {arXiv},
	author = {Kong, Liang and Wen, Xiao-Gang and Zheng, Hao},
	eprint = {1502.01690},
	title = {Boundary-bulk relation for topological orders as the functor mapping higher categories to their centers},
	year = {2015}}

@article{Kong2020,
	author = {Kong, Liang and Lan, Tian and Wen, Xiao-Gang and Zhang, Zhi-Hao and Zheng, Hao},
	doi = {10.1103/PhysRevResearch.2.043086},
	issue = {4},
	journal = {Phys. Rev. Res.},
	month = {Oct},
	numpages = {53},
	pages = {043086},
	publisher = {American Physical Society},
	title = {Algebraic higher symmetry and categorical symmetry: A holographic and entanglement view of symmetry},
	url = {https://link.aps.org/doi/10.1103/PhysRevResearch.2.043086},
	volume = {2},
	year = {2020}}

@article{Bhardwaj2020,
	author = {Bhardwaj, Lakshya and Lee, Yasunori and Tachikawa, Yuji},
	doi = {10.1007/jhep11(2020)141},
	issn = {1029-8479},
	journal = {Journal of High Energy Physics},
	month = nov,
	number = {11},
	publisher = {Springer Science and Business Media LLC},
	title = {SL(2, Z) action on QFTs with Z2 symmetry and the Brown-Kervaire invariants},
	url = {http://dx.doi.org/10.1007/JHEP11(2020)141},
	volume = {2020},
	year = {2020}}

@article{Pulmann2021,
	author = {Pulmann, J{\'a}n and {\v S}evera, Pavol and Valach, Fridrich},
	doi = {10.4310/atmp.2021.v25.n1.a5},
	issn = {1095-0753},
	journal = {Advances in Theoretical and Mathematical Physics},
	number = {1},
	pages = {241--274},
	publisher = {International Press of Boston},
	title = {A nonabelian duality for (higher) gauge theories},
	url = {http://dx.doi.org/10.4310/ATMP.2021.v25.n1.a5},
	volume = {25},
	year = {2021}}

@article{Gaiotto2021,
	author = {Gaiotto, Davide and Kulp, Justin},
	doi = {10.1007/jhep02(2021)132},
	issn = {1029-8479},
	journal = {Journal of High Energy Physics},
	month = feb,
	number = {2},
	publisher = {Springer Science and Business Media LLC},
	title = {Orbifold groupoids},
	url = {http://dx.doi.org/10.1007/JHEP02(2021)132},
	volume = {2021},
	year = {2021}}

@article{Lichtman2021,
	author = {Lichtman, Tsuf and Thorngren, Ryan and Lindner, Netanel H. and Stern, Ady and Berg, Erez},
	doi = {10.1103/PhysRevB.104.075141},
	issue = {7},
	journal = {Phys. Rev. B},
	month = {Aug},
	numpages = {27},
	pages = {075141},
	publisher = {American Physical Society},
	title = {Bulk anyons as edge symmetries: Boundary phase diagrams of topologically ordered states},
	url = {https://link.aps.org/doi/10.1103/PhysRevB.104.075141},
	volume = {104},
	year = {2021}}

@article{kong2020mathematical,
	author = {Kong, Liang and Zheng, Hao},
	journal = {Journal of High Energy Physics},
	number = {2},
	pages = {1--62},
	publisher = {Springer},
	title = {A mathematical theory of gapless edges of 2d topological orders. Part I},
	volume = {2020},
	year = {2020}}

@article{Kong2020classification,
	author = {Kong, Liang and Lan, Tian and Wen, Xiao-Gang and Zhang, Zhi-Hao and Zheng, Hao},
	doi = {10.1007/jhep09(2020)093},
	issn = {1029-8479},
	journal = {Journal of High Energy Physics},
	month = sep,
	number = {9},
	publisher = {Springer Science and Business Media LLC},
	title = {Classification of topological phases with finite internal symmetries in all dimensions},
	url = {http://dx.doi.org/10.1007/JHEP09(2020)093},
	volume = {2020},
	year = {2020}}

@article{Apruzzi2023,
	author = {Apruzzi, Fabio and Bonetti, Federico and Garc{\'\i}a Etxebarria, I{\~n}aki and Hosseini, Saghar S. and Sch{\"a}fer-Nameki, Sakura},
	doi = {10.1007/s00220-023-04737-2},
	issn = {1432-0916},
	journal = {Communications in Mathematical Physics},
	month = may,
	number = {1},
	pages = {895--949},
	publisher = {Springer Science and Business Media LLC},
	title = {Symmetry TFTs from String Theory},
	url = {http://dx.doi.org/10.1007/s00220-023-04737-2},
	volume = {402},
	year = {2023}}

@article{Chatterjee2023symmetry,
	archiveprefix = {arXiv},
	author = {Chatterjee, Arkya and Wen, Xiao-Gang},
	doi = {10.1103/PhysRevB.107.155136},
	eprint = {2203.03596},
	issue = {15},
	journal = {Phys. Rev. B},
	month = {Apr},
	numpages = {34},
	pages = {155136},
	publisher = {American Physical Society},
	title = {Symmetry as a shadow of topological order and a derivation of topological holographic principle},
	url = {https://link.aps.org/doi/10.1103/PhysRevB.107.155136},
	volume = {107},
	year = {2023}}

@article{Moradi2023,
	author = {Moradi, Heidar and Moosavian, Seyed Faroogh and Tiwari, Apoorv},
	doi = {10.21468/SciPostPhysCore.6.4.066},
	journal = {SciPost Phys. Core},
	pages = {066},
	publisher = {SciPost},
	title = {Topological holography: Towards a unification of Landau and beyond-Landau physics},
	url = {https://scipost.org/10.21468/SciPostPhysCore.6.4.066},
	volume = {6},
	year = {2023}}

@misc{Freed2023topological,
	archiveprefix = {arXiv},
	author = {Daniel S. Freed and Gregory W. Moore and Constantin Teleman},
	eprint = {2209.07471},
	primaryclass = {hep-th},
	title = {Topological symmetry in quantum field theory},
	year = {2023}}

@article{Kong2018,
	author = {Kong, Liang and Zheng, Hao},
	journal = {Nuclear Physics B},
	pages = {140--165},
	publisher = {Elsevier},
	title = {Gapless edges of 2d topological orders and enriched monoidal categories},
	volume = {927},
	year = {2018}}

@article{Kong2021,
	author = {Kong, Liang and Zheng, Hao},
	journal = {Nuclear Physics B},
	pages = {115384},
	publisher = {Elsevier},
	title = {A mathematical theory of gapless edges of 2d topological orders. Part II},
	volume = {966},
	year = {2021}}

@article{Kong2022one,
	author = {Kong, Liang and Wen, Xiao-Gang and Zheng, Hao},
	journal = {Journal of High Energy Physics},
	number = {3},
	pages = {1--32},
	publisher = {Springer},
	title = {One dimensional gapped quantum phases and enriched fusion categories},
	volume = {2022},
	year = {2022}}

@article{Kong2022categories,
	author = {Kong, Liang and Zheng, Hao},
	journal = {Journal of High Energy Physics},
	number = {8},
	pages = {1--44},
	publisher = {Springer},
	title = {Categories of quantum liquids I},
	volume = {2022},
	year = {2022}}

@article{kong2024categories,
	author = {Kong, Liang and Zheng, Hao},
	journal = {Communications in Mathematical Physics},
	number = {9},
	pages = {203},
	publisher = {Springer},
	title = {Categories of quantum liquids II},
	volume = {405},
	year = {2024}}

@article{Xu2024,
	author = {Xu, Rongge and Zhang, Zhi-Hao},
	journal = {Physical Review B},
	number = {15},
	pages = {155106},
	publisher = {APS},
	title = {Categorical descriptions of one-dimensional gapped phases with Abelian onsite symmetries},
	volume = {110},
	year = {2024}}

@article{Quspin2017,
	title = {QuSpin: a Python package for dynamics and exact diagonalisation of quantum many body systems part I: spin chains},
	pages = {003},
	author = {Weinberg, Phillip and Bukov, Marin},
	journal = {SciPost Phys.},
	volume = {2},
	year = {2017},
	publisher = {SciPost},
	doi = {10.21468/SciPostPhys.2.1.003},
	url = {https://scipost.org/10.21468/SciPostPhys.2.1.003}
}

@article{Quspin2019,
	title = {QuSpin: a Python package for dynamics and exact diagonalisation of quantum many body systems. Part II: bosons, fermions and higher spins},
	pages = {020},
	author = {Weinberg, Phillip and Bukov, Marin},
	journal = {SciPost Phys.},
	volume = {7},
	year = {2019},
	publisher = {SciPost},
	doi = {10.21468/SciPostPhys.7.2.020},
	url = {https://scipost.org/10.21468/SciPostPhys.7.2.020}
}

@article{Kitaev2003,
  author = {Kitaev, A Yu},
  journal = {Annals of physics},
  number = {1},
  pages = {2--30},
  publisher = {Elsevier},
  title = {Fault-tolerant quantum computation by anyons},
  volume = {303},
  year = {2003},
}

@article{Beigi2011,
  author = {Beigi, Salman and Shor, Peter W and Whalen, Daniel},
  title = {The quantum double model with boundary: condensations and symmetries},
  journal = {Communications in mathematical physics},
  year = {2011},
  pages = {663--694},
  publisher = {Springer},
  volume = {306},
}

@article{Chepiga2019,
  author = {Chepiga, Natalia and Mila, Fr{\'e}d{\'e}ric},
  title = {Floating Phase versus Chiral Transition in a 1D Hard-Boson Model},
  journal = {Phys. Rev. Lett.},
  year = {2019},
  volume = {122},
  issue = {1},
  pages = {017205},
  numpages = {5},
  month = {Jan},
  publisher = {American Physical Society},
  doi = {10.1103/PhysRevLett.122.017205},
  url = {https://link.aps.org/doi/10.1103/PhysRevLett.122.017205},
}

@article{Maceira2022,
  author = {Maceira, Ivo A. and Chepiga, Natalia and Mila, Fr{\'e}d{\'e}ric},
  title = {Conformal and chiral phase transitions in Rydberg chains},
  journal = {Phys. Rev. Res.},
  year = {2022},
  volume = {4},
  issue = {4},
  pages = {043102},
  numpages = {10},
  month = {Nov},
  publisher = {American Physical Society},
  doi = {10.1103/PhysRevResearch.4.043102},
  url = {https://link.aps.org/doi/10.1103/PhysRevResearch.4.043102},
}

@article{Iqbal2022,
  author = {Nabil Iqbal and John McGreevy},
  doi = {10.21468/SciPostPhys.13.5.114},
  journal = {SciPost Phys.},
  pages = {114},
  publisher = {SciPost},
  title = {{Mean string field theory: Landau-Ginzburg theory for 1-form symmetries}},
  url = {https://scipost.org/10.21468/SciPostPhys.13.5.114},
  volume = {13},
  year = {2022},
}

@article{Johnson-Freyd2022,
  author = {Johnson-Freyd, Theo},
  date = {2022/07/01},
  doi = {10.1007/s00220-022-04380-3},
  id = {Johnson-Freyd2022},
  isbn = {1432-0916},
  journal = {Communications in Mathematical Physics},
  number = {2},
  pages = {989--1033},
  title = {On the Classification of Topological Orders},
  url = {https://doi.org/10.1007/s00220-022-04380-3},
  volume = {393},
  year = {2022},
}

@misc{Kong2014arxiv,
  archiveprefix = {arXiv},
  author = {Liang Kong and Xiao-Gang Wen},
  eprint = {1405.5858},
  primaryclass = {cond-mat.str-el},
  title = {Braided fusion categories, gravitational anomalies, and the mathematical framework for topological orders in any dimensions},
  year = {2014},
}

@article{Alford1990,
  author = {Alford, M. G. and Benson, Katherine and Coleman, Sidney and March-Russell, John and Wilczek, Frank},
  doi = {10.1103/PhysRevLett.64.1632},
  issue = {14},
  journal = {Phys. Rev. Lett.},
  month = {Apr},
  numpages = {0},
  pages = {1632--1635},
  publisher = {American Physical Society},
  title = {Interactions and excitations of non-Abelian vortices},
  url = {https://link.aps.org/doi/10.1103/PhysRevLett.64.1632},
  volume = {64},
  year = {1990},
}

@article{Preskill1990,
  author = {Preskill, John and Krauss, Lawrence M.},
  date = {1990/09/03/},
  doi = {https://doi.org/10.1016/0550-3213(90)90262-C},
  isbn = {0550-3213},
  journal = {Nuclear Physics B},
  number = {1},
  pages = {50--100},
  title = {Local discrete symmetry and quantum-mechanical hair},
  url = {https://www.sciencedirect.com/science/article/pii/055032139090262C},
  volume = {341},
  year = {1990},
}

@article{Tantivasadakarn2024,
  author = {Tantivasadakarn, Nathanan and Chen, Xie},
  doi = {10.1103/PhysRevB.109.165149},
  issue = {16},
  journal = {Phys. Rev. B},
  month = {Apr},
  numpages = {8},
  pages = {165149},
  publisher = {American Physical Society},
  title = {String operators for Cheshire strings in topological phases},
  url = {https://link.aps.org/doi/10.1103/PhysRevB.109.165149},
  volume = {109},
  year = {2024},
}

\end{document}